\pdfoutput=1
\documentclass[graybox]{svmult}
\usepackage{mathptmx}       
\usepackage{helvet}         
\usepackage{courier}        
\usepackage{type1cm}        
\usepackage{makeidx}         
\usepackage{graphicx}        
\usepackage{multicol}        
\usepackage[bottom]{footmisc}

\usepackage{amssymb,mathtools}
\usepackage{lineno}
\usepackage{url, csvsimple}
\usepackage{verbatim,  booktabs}
\usepackage{algorithmic, algorithm2e}
\usepackage{soul}
\usepackage{cite}  

\newtheorem{dfn}[table]{Definition}
\newtheorem{thm}[table]{Theorem}
\newtheorem{lem}[table]{Lemma}

\newtheorem{cor}[table]{Corollary}

\newcommand{\R}{\mathbb R}
\newcommand{\al}{\alpha}
\newcommand{\be}{\beta}
\newcommand{\ga}{\gamma}
\newcommand{\de}{\delta}

\newcommand{\ep}{\varepsilon}

\newcommand{\MST}{\mathrm{MST}}
\newcommand{\VR}{\mathrm{VR}}

\newcommand{\ind}{\mathrm{ind}}
\newcommand{\opt}{\mathrm{opt}}
\newcommand{\core}{\mathrm{core}}
\newcommand{\sk}{\mathrm{ASk}}
\newcommand{\dep}{\mathrm{depth}}

\newcommand{\myskip}{\smallskip}

\makeindex             

\begin{document}
\title*{A fast approximate skeleton with guarantees for any cloud of points in a Euclidean space}
\titlerunning{A fast approximate skeleton for any cloud} 
\author{Yury Elkin, Di Liu, Vitaliy Kurlin}
\institute{Materials Innovation Factory and Computer Science department, University of Liverpool, Liverpool L69 3BX, UK.
\email{vitaliy.kurlin@gmail.com, http://kurlin.org} }
%
%
\maketitle
\abstract{
The tree reconstruction problem is to find an embedded straight-line tree that approximates a given cloud of unorganized points in $\mathbb{R}^m$ up to a certain error. 
A practical solution to this problem will accelerate a discovery of new colloidal products with desired physical properties such as viscosity.
We define the Approximate Skeleton of any finite point cloud $C$ in a Euclidean space with theoretical guarantees.
The Approximate Skeleton $\sk(C)$ always belongs to a given offset of $C$, i.e. the maximum distance from $C$ to $\sk(C)$ can be a given maximum error.
The number of vertices in the Approximate Skeleton is close to the minimum number in an optimal tree by factor 2.
The new Approximate Skeleton of any unorganized point cloud $C$ is computed in a near linear time in the number of points in $C$.
Finally, the Approximate Skeleton outperforms past skeletonization algorithms on the size and accuracy of reconstruction for a large dataset of real micelles and random clouds. 
}

%


\section{Introduction: reconstructions from unorganized clouds}
\label{sec:intro}

\begin{figure}[h]
\includegraphics[width=0.48\linewidth]{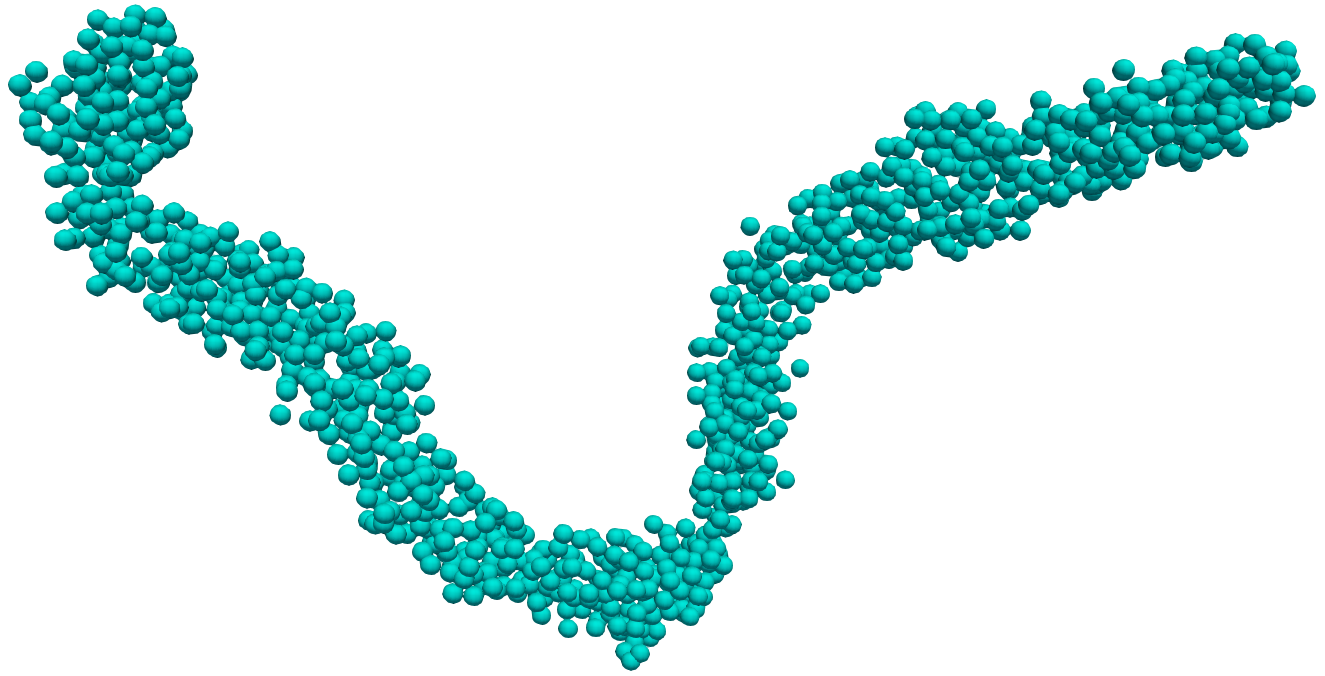}
\includegraphics[width=0.48\linewidth]{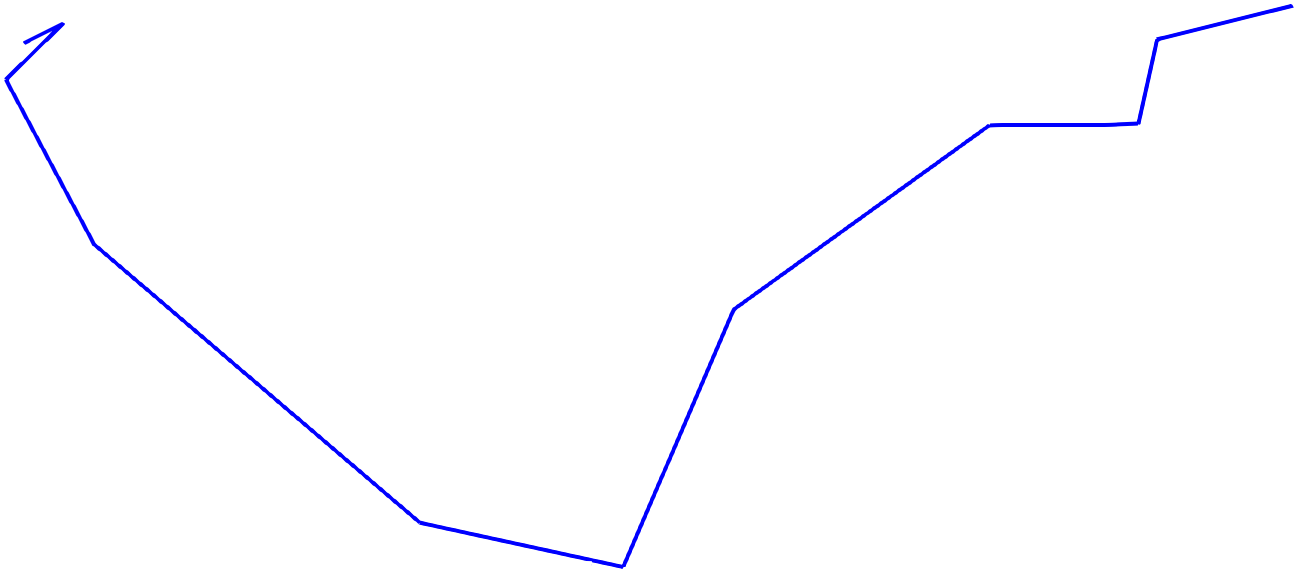}

\includegraphics[width=0.48\linewidth]{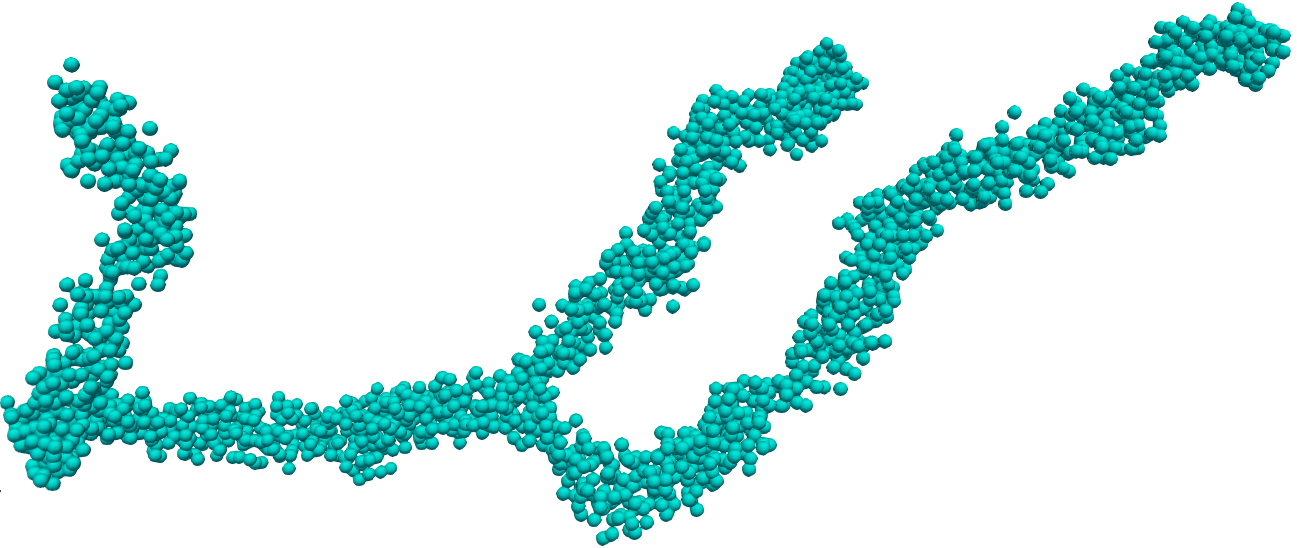}
\includegraphics[width=0.48\linewidth]{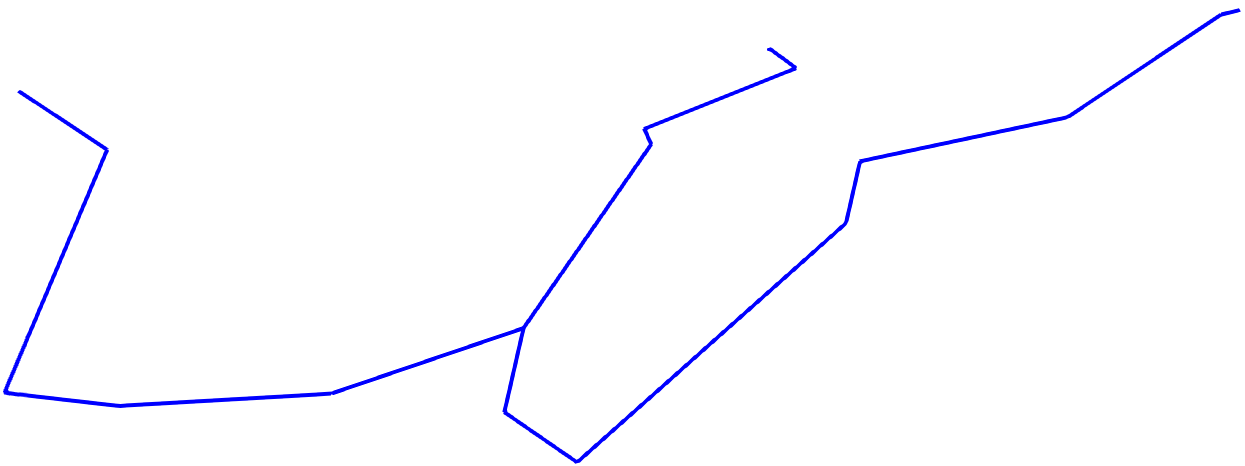}
\caption{
\textbf{Left}: point clouds $C$ from real micelles. 
\textbf{Right}: Approximate Skeletons $\sk(C)$.}
\label{fig:cylindrical+branched}
\end{figure}

Potential molecules for new colloidal products are tested by simulations that produce unorganized finite clouds of points (one point per molecule in Fig.~\ref{fig:cylindrical+branched}).
Molecules tend to form clusters (called {\em micelles}) whose shapes (degrees of branching, edge-lengths) affect physical properties of colloidal products, e.g. their viscosity.

These 3D micelles can have complicated branched shapes as in Fig.~\ref{fig:Christmas_outputs} and are visually analyzed by human experts who struggle to make reliable measurements quickly.
To substantially speed-up the discovery of new molecules, we propose a new Approximate Skeleton $\sk(C)$ to solve the following problem.
\smallskip

\noindent
\textbf{The tree reconstruction problem}.
Given a point cloud $C\subset\R^m$ and an error $\ep$, design a fast algorithm to build a straight-line tree $T\subset\R^m$ (see Definition~\ref{dfn:graph}) that has a minimum number of vertices and whose $\ep$-offset (neighborhood) covers $C$.
\medskip

The first (combinatorial) guarantee is for the number of vertices in $\sk(C)$, which is close to the minimum number in an optimal tree for a given approximation error by factor 2, see Theorem~\ref{thm:size}.
The second (geometric) guarantee about a near linear time for building $\sk(C)$ is the number of points $n$ in $C$, see Corollary~\ref{cor:time}.
\myskip

To automatically characterize branching shapes of micelles (clusters of molecules in colloids), an Approximate Skeleton $\sk(C)$ allows us to compute
\smallskip

\noindent
$\bullet$
the topological type of any unorganized cloud $C$, e.g. count all {\em non-trivial} vertices of $\sk(C)\subset\R^m$ whose degree is 1 (endpoints) or more than 2 (branching);
\smallskip

\noindent
$\bullet$
the geometric characteristics of $C$, e.g. edge-lengths of $\sk(C)$;
\smallskip

\noindent
$\bullet$
the error of approximating a cloud $C$ by its skeleton $\sk(C)$, see  Table~\ref{tab:micelles}.  
\myskip

Here is the pipeline of the Approximate Skeleton $\sk(C)$.
\smallskip

\noindent
Stage 1 in section~\ref{sec:core}: for a cloud $C\subset\R^m$, we build an initial tree $\core(C)$, which has a small number of branching vertices within a Minimum Spanning Tree of $C$.
\smallskip

\noindent
Stage 2 in section~\ref{sec:ask}: replace polygonal paths of $\core(C)$ by approximate paths with much fewer vertices to get $\sk(C)$ in a near linear time within a given error.

\begin{figure}[h]
\includegraphics[width=1.0\linewidth]{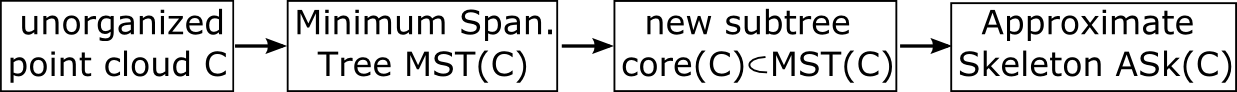}
\caption{Pipeline to compute an Approximate Skeleton $\sk(C)$ :
$\MST(C)$ is classical, the new subtree $\core(C)\subset\MST(C)$ is introduced in Definition~6 in section~3, final $\sk(C)$ is built in section~4}.
\label{fig:pipeline}
\end{figure}

\noindent
The key novelty and contributions to the data skeletonization are the following.
\smallskip

\noindent
$\bullet$
Theorem~\ref{thm:size} guarantees a small number of vertices in the Approximate Skeleton $\sk(C)$ close to the minimum by factor 2 in an optimal tree within a given error.
\smallskip

\noindent
$\bullet$
Corollary~\ref{cor:time} guarantees a near linear time  to compute $\sk(C)$ within an error.

\section{Basic definitions and a review of the related past work}
\label{sec:review}

\begin{dfn}[a straight-line graph, $\ep$-approximation]
\label{dfn:graph}
A {\em straight-line} graph $G\subset\R^m$ (briefly, a {\em graph}) consists of vertices at points $q_1,\dots,q_k\in\R^m$ and undirected straight-line edges connecting pairs $q_i,q_j$, $i\neq j$, in such a way that any edges meet only at their common vertex.
Let $d$ be the Euclidean distance.
For $\ep>0$, a cloud $C\subset\R^m$ is $\ep$-{\em approximated} by a graph $G$ if $C$ is within the $\ep$-{\em offset} that is the union of $\ep$-balls at all points of $G$, i.e.
$G^{\ep}=\{p\in\R^m \;|\; d(p,q)\leq\ep \mbox{ for some } q\in G\}$.
\end{dfn}

\noindent
\textbf{Past algorithms without guarantees}.
Singh et al. \cite{SCP00} approximated a cloud $C\subset\R^m$ by a subgraph of a Delaunay triangulation, which requires $O(n^{\lceil{m/2}\rceil})$ time for $n$ points of $C$
and the three thresholds: a minimum number $K$ of edges in a cycle and $\de_{min},\de_{max}$ for inserting/merging 2nd order Voronoi regions.
Similar parameters are need for {\em principal curves} \cite{KK02}, which were later extended to iteratively computed {\em elastic maps} \cite{GZ09}.
Since it is often hard to estimate a rate of convergence for iterative algorithms, we discuss below non-iterative methods with theoretical guarantees.
\medskip

\noindent
\textbf{The metric graph reconstruction} (MGR) takes as an input a large metric graph $Y$, which is an abstract graph with weighted edges  and outputs a smaller abstract metric graph $\hat X$.
The distance between any points of a metric graph is defined as the length of a shortest path these points.
If $Y$ is a good $\ep$-approximation to an unknown graph $X$, then M.~Aanjaneya et al. \cite[Theorem~5]{ACCGGM12} proved the existence of a homeomorphism $X\to\hat X$ that distorts the metrics on $X$ and $\hat X$ with a multiplicative factor $1+c\ep$ for $c>\frac{30}{b}$, where $b>14.5\ep$ is the length of a shortest edge of $X$.
\smallskip

\noindent
The authors of the Reeb graph skeletonization \cite[page~3]{GSBW11} have checked that for the MGR algorithm from \cite{ACCGGM12} ``it is often hard to find suitable parameters in practice, and such local decisions tend to be less reliable when the input data are not as nice (such as a `fat' junction region)", see this junction in the 2nd picture of Fig.~\ref{fig:cylindrical+branched}.

\begin{dfn}[a Reeb graph]
\label{dfn:Reeb}
Given a topological space $K\subset\R^m$ (or ) with a function $f:K\to\R$, the {\em Reeb} graph $R_f(K)$ is obtained from $K$ by collapsing each connected components of every level set of $f$ to a single point, so the {\em Reeb} graph $R_f(K)$ is the quotient of $K$ by the equivalence relation $a\sim b$ if and only if $f(a)=t=f(b)$ and the points $a,b\in K$ are in the same connected component of $f^{-1}(t)\subset K$.
\end{dfn}

\noindent
\textbf{Skeletonization via Reeb-type graphs}.
The {\em Vietoris-Rips} complex $\VR(C;\al)$ on a cloud $C$ consists of all simplices spanned by points whose pairwise distances are at most $\al$.
Starting from a noisy sample $C$ of an unknown graph $G$ with a scale parameter, X.~Ge et al. \cite[Theorem~3.1]{GSBW11} proved that the Reeb graph of $\VR(C;\al)$ has a correct homotopy type if there is a triangulated space $K$ with a continuous deformation $h:K\to G$ that $\ep$-approximates the metrics of $K,G$.
The {\em homotopy type} of a graph is the equivalence class of graphs under deformations when any edge (with distinct endpoints) can be collapsed to a point.
\medskip

\noindent
\textbf{The graph reconstruction by discrete Morse theory} (DMT).
T.~Dey et al. \cite{dey2018graph} substantially improved the discrete Morse-based framework \cite{delgado2015skeletonization} and proved new homotopy guarantees when an input is a density function $\rho:K\to\R$, which `concentrates' around a hidden geometric graph $G$.
The key advantage of this approach is the unbounded noise model that allows outliers far away from the underlying graph $G$, which has found practical applications to map reconstructions \cite{wang2015efficient, dey2017improved}.
\medskip

Since the molecules of a micelle form an unorganized cloud of points (with large bounded noise) around hidden tree structures, the Tree Reconstruction problem in section~\ref{sec:intro} essentially differs from the above approaches.
An initial unorganized cloud of points is not an abstract metric graph (as in the metric graph reconstruction problem) and not a simplicial complex with scalar values at vertices (as in the discrete Morse theory approach), so extra pre-processing was needed in section~\ref{sec:experiments}.
\medskip

\noindent
\textbf{The $\al$-Reeb graph $G$} by F.~Chazal et al. \cite{CHS15} solves the metric graph reconstruction problem, where the input is not an unorganized cloud, but a large metric graph $X$ that should be approximated by a smaller graph $\hat X$.
For a base point $p\in X$, the image of the distance function $d(p,*):X\to\R$ is covered by intervals $I_j$ having a length $\al$ and 50\% overlap.
Every connected component of $f^{-1}(I_j)\subset X$ defines a node in the $\al$-Reeb graph $G$.
Two nodes are linked if the corresponding components overlap. 
\smallskip

\noindent
Informally, $\al$ controls the size of a subset of $X$ that maps to a single vertex of $G$.
Theorem~4.9 in \cite{CHS15} says that if $X$ is $\ep$-close to an unknown graph with edges of minimum length $8\ep$, the output $G$ is $34(\be(G)+1)\ep$-close to $X$ in the Gromov-Hausdorff distance between spaces, not within one space, where $\be(G)$ is the first Betti number of $G$. 
The algorithm has the fast time $O(n\log n)$ for $n$ points in $X$.
Similarly to Reeb graphs, $\al$-Reeb graphs are abstract without an intrinsic embedding into the space of the cloud $C$ and can have self-intersections even for $X\subset\R^2$.
\medskip

\noindent
\textbf{The Mapper} \cite{SMG07} extends any clustering algorithm and outputs a network of interlinked clusters and needs a user-defined function $f:C\to\R$, which helps to link different clusters of a cloud $C$.
Another parameter is a covering of the image of $f$ by a given number $k$ of intervals $I_j$ (often with 50\% overlap).
Each of $k$ subclouds $f^{-1}(I_j)\subset C$ is clustered.
Every cluster defines a node in the Mapper graph.
Two nodes are linked if the corresponding clusters overlap.
M.~Carri\'ere et al. \cite{CS17} have proved first theoretical guarantees for the Mapper output.
\medskip

More recent persistence-based algorithms for graph reconstruction \cite{kurlin2015homologically, kurlin2015one, kalisnik2019higher} and image segmentation \cite{forsythe2016resolution, forsythe2017convex, kurlin2017superpixels, kurlin2020persistence} essentially find most persistent cycles hidden in a cloud, hence go beyond the tree reconstruction problem in section~\ref{sec:intro}.
\medskip

\noindent
\textbf{Straightening polygonal curves} is a key ingredient in many skeletonization algorithms.
Douglas-Peucker's heuristic \cite{DP73} approximates a long zigzag line by a simpler line with fewer vertices, see section~\ref{sec:ask}.
The elegant algorithm by P.~Agarwal et al. \cite{AHMW05} guarantees a near linear time and a small number of vertices in a final polygonal approximation when used with the Frechet distance between curves in $\R^2$.
For the Hausdorff distance and  higher dimensions, there is no near linear time straightening with guarantees on the size of a skeleton to our best knowledge. 

\begin{dfn}[$\MST(C)$]
\label{dfn:MST}
For a cloud $C\subset\R^m$, a {\em Minimum Spanning Tree} $\MST(C)$ is a connected graph that has (1) the vertex set $C$, (2) no cycles, and (3) a minimum total length, where lengths of edges are measured in the Euclidean distance.
\end{dfn}

If all distances between points of $C$ are distinct, then $\MST(C)$ is unique.
We write a Minimum Spanning Tree, similarly an Approximate Skeleton, to cover all cases.

\begin{thm}
\label{thm:MST_time}
\cite[Theorem~5.1]{EMST} 
For any cloud $C\subset\R^m$ of $n$ points, a Minimum Spanning Tree $\MST(C)$ can be computed in time $O(\max\{c^6,c_p^2c^2_l) \}c^{10}n\log n\,\al(n))$, where 
 $\al(n)$ is the inverse Ackermann function; 
 $c,c_p,c_l$ are defined in Appendix~A.
\end{thm}

\section{A new tree $\core(C)$ defined for any point cloud $C\subset\R^m$}
\label{sec:core}

This section introduces an important subtree $\core(C)\subset\MST(C)$, which has many fewer non-trivial vertices than a usually `hairy' $\MST(C)$ from Definition~\ref{dfn:MST}.
\myskip

A tree $\core(C)$ might still have too many zigzags and will be replaced by a better tree $\sk(C)$ with fewer vertices in section~\ref{sec:ask}.
A vertex of a degree $k\neq 2$ is called (topologically) {\em non-trivial}, because any vertex of degree~2 can be potentially removed by straightening algorithms in section~\ref{sec:ask}.
Since $\MST(C)$ contains many non-trivial vertices, the next hard step is to identify those few vertices of $\MST(C)$ that represent `true' vertices of a tree $T$, which we try to reconstruct from $C$.
\medskip

\begin{figure}
\includegraphics[width=0.48\linewidth]{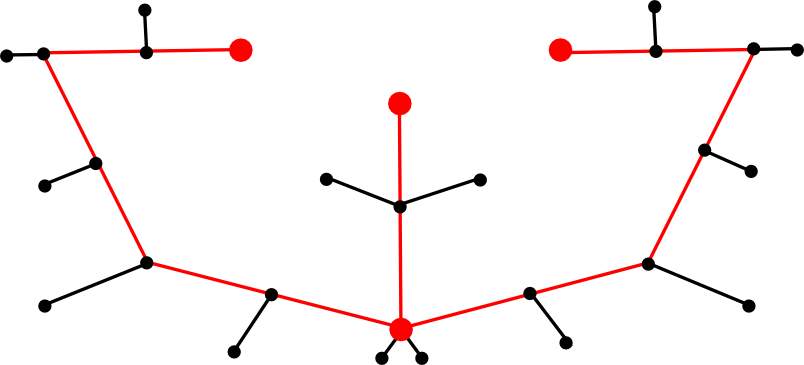}
\hspace*{2mm}
\includegraphics[width=0.48\linewidth]{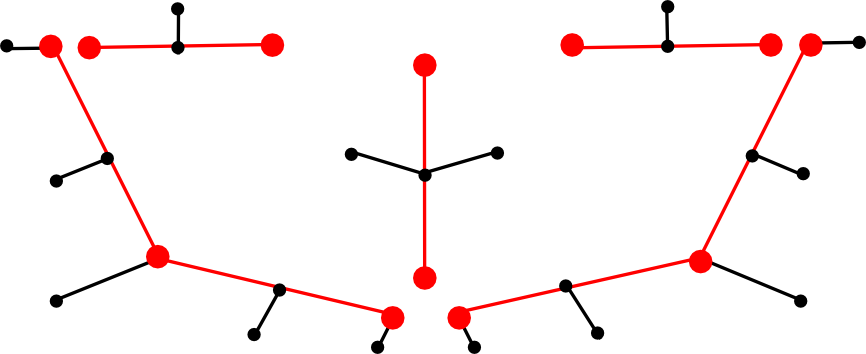}
\caption{\textbf{Left}. One vertex (large red dot at the bottom) of $\MST(C)$ has a high depth by Definition~\ref{dfn:depth} and is connected by longest paths to 3 vertices of degree~1.
The other vertices have at most 2 disjoint long paths within $\MST(C)$.
\textbf{Right}. The red monotone paths of $\core(C)$ and subclouds from Algorithm~\ref{alg:subclouds} are shown disjointly.}
\label{fig:depth}
\end{figure}

Definition~\ref{dfn:depth} introduces the depth characterizing how deep a vertex sits within $\MST(C)$. 
At a deep vertex of a degree $k\geq 3$ at least 3 sufficiently long paths (without common edges) should meet, see the 3 red long paths in Fig.~\ref{fig:depth}.
The previous procedural approach by M.~Aanjaneya et al. \cite[Fig.~1b]{ACCGGM12} to detect branching points in a shape of $C$ used more parameters than a single branching factor $\be$ below.

\begin{dfn}[deep vertices]
\label{dfn:depth}
For a cloud $C\subset\R^m$ and a vertex $v\in\MST(C)$ of a degree $k\geq 3$, let $B_1,\dots,B_k\subset\MST(C)$ be the branches (subtrees) joined at the vertex $v$.
Let $l_i$ be the length of a longest path within the branch $B_i$ from $v$ to another vertex, $i=1,\dots,k$.
Assuming that $l_1\geq l_2\geq\dots$, set $\dep(v)=\min\{l_1,l_2,l_3\}$.
Let $l(C)$ be the average edge-length of $\MST(C)$.
For a branching factor $\be>0$, the vertices of $\MST(C)$ whose depths are larger than $\be l(C)$ are called {\em deep}.
\end{dfn}

Taking the minimum $\dep(v)=\min\{l_1,l_2,l_3\}$ above guarantees that vertices in any short branches of $\MST(C)$ are not deep, hence deep vertices can not form small cliques.
The experiments on real micelles in section~\ref{sec:experiments} justify that Definition~\ref{dfn:depth} separates deep vertices from other shallow vertices for a long range of the factor $\be$.
\smallskip

\begin{figure}[h]
\includegraphics[width=0.48\linewidth]{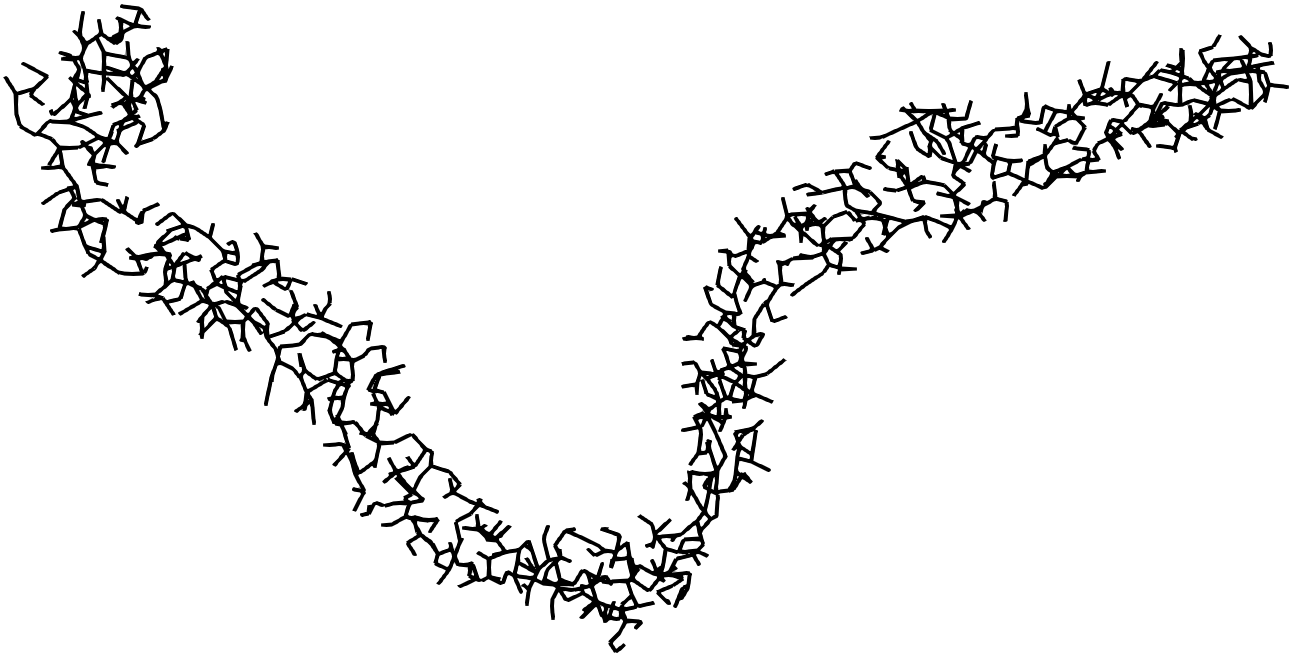}
\includegraphics[width=0.48\linewidth]{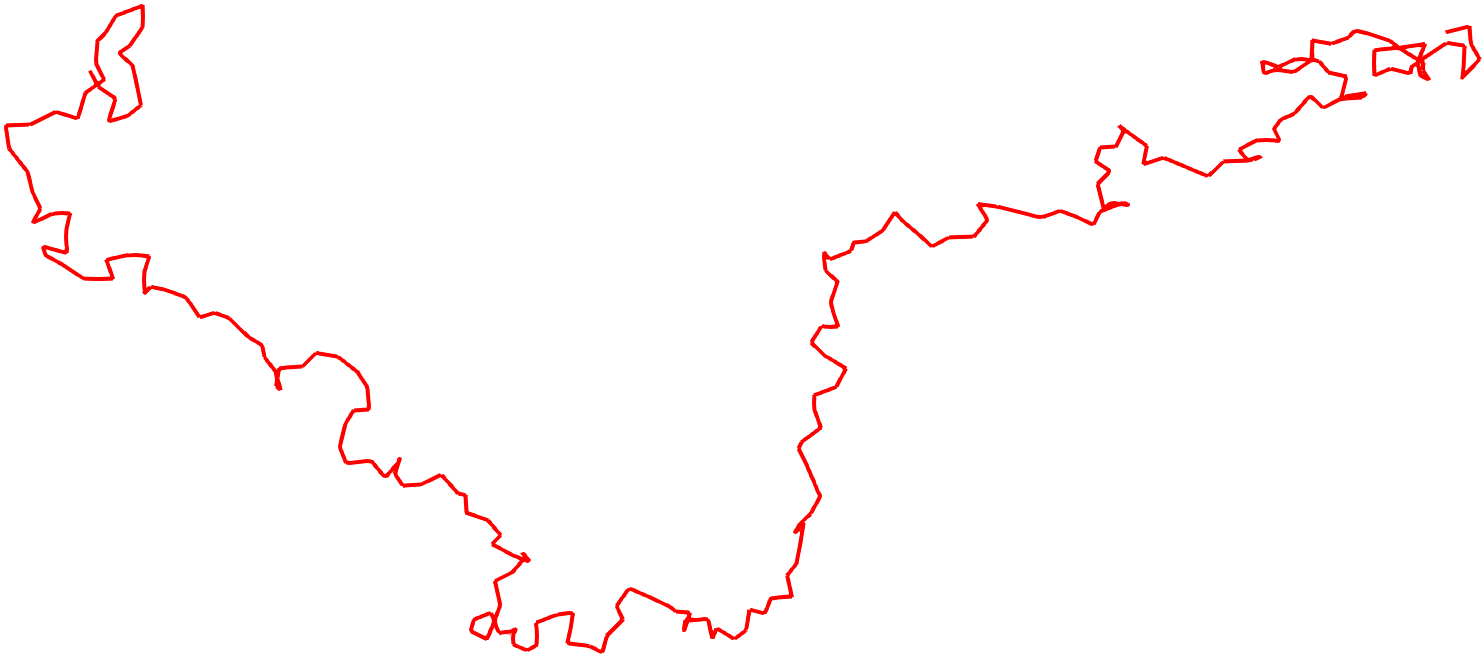}

\includegraphics[width=0.48\linewidth]{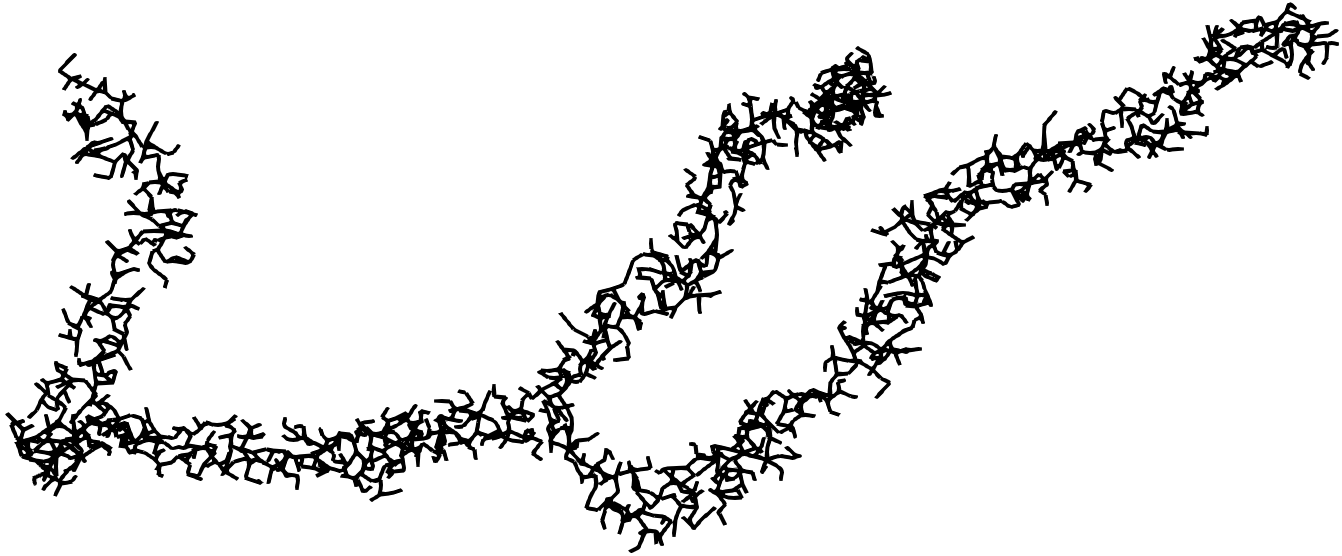}
\includegraphics[width=0.48\linewidth]{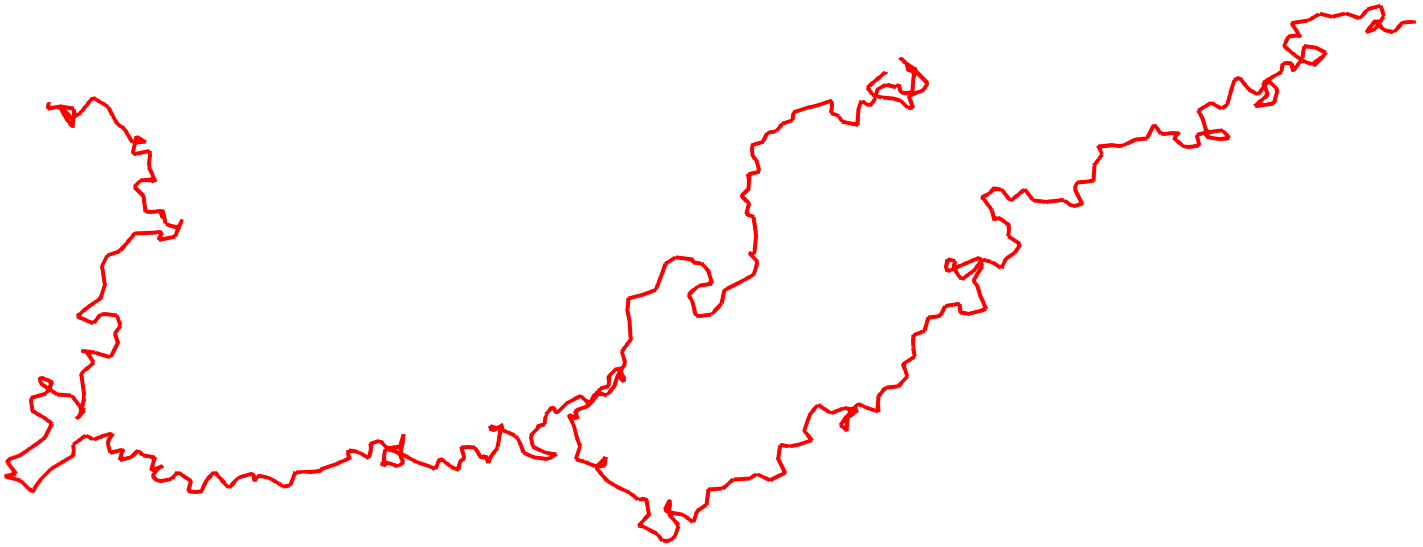}
\caption{\textbf{Left}: black $\MST(C)$ for the two clouds $C$ in Fig.~\ref{fig:cylindrical+branched}. \textbf{Right}: red $\core(C)$ in Definition~\ref{dfn:core}.}
\label{fig:MST+core}
\end{figure}

Definition~\ref{dfn:core} introduces a subtree $\core(C)$, which non-essentially depends on the branching factor $\be$ and better approximates a cloud $C$ than $\MST(C)$, see Fig.~\ref{fig:depth}.

\begin{dfn}[$\core(C)$]
\label{dfn:core}
In Definition~\ref{dfn:depth}, if we remove all deep vertices $v_1,\dots,v_m$, $\MST(C)$ splits into several subtrees.
If the closure of such a subtree $S$ has two deep vertices $v_i,v_j$, they are joined by a unique path $P_{ij}\subset S$.
If $S$ has one deep vertex $v_i$, take a longest path $P_i\subset S$ from $v_i$ to another vertex $v'_i\in S$.
We ignore $P_i$ if its length is less than $\be l(C)$, where $\be$ is the branching factor from Definition~\ref{dfn:depth}.
All the vertices $v_i,v'_i$ and the paths $P_{ij},P_i$ between them form the subtree $\core(C)\subset\MST(C)$. 
\end{dfn}

If the closure of a subtree $S$ above has $k\geq 3$ deep vertices $v_1,v_2,v_3$, then $S$ contains a vertex $v$ with at least 3 paths to $v_1,v_2,v_3$.
Then $\dep(v)>\dep(v_i)$, $i=1,2,3$, so $v$ is also deep and $S$ should be split by removing $v$.
Hence $k\leq 2$. 
\smallskip

In Fig.~\ref{fig:depth} the two black edges at the red deep vertex $v$ of degree~5 are too short, hence ignored in Definition~\ref{dfn:core}.
The tree $\core(C)$ consists of only 3 red long paths meeting at $v$.
Here are the steps of Stage 1 for the Approximate Skeleton $\sk(C)$.
\smallskip

\noindent
\textbf{Step 1a}. 
If needed, split a cloud $C$ in clusters to approximate them below.
\smallskip

\noindent
\textbf{Step 1b}. 
If $C$ is one cluster, find $\MST(C)$ by the fast algorithm from Theorem~\ref{thm:MST_time}.
\smallskip

\noindent
\textbf{Step 1c}. 
Find the depths of vertices in $\MST(C)$ by Algorithm~\ref{alg:depths} in Appendix~A. 
\smallskip

\noindent
\textbf{Step 1d}. 
Identify all deep vertices of $\MST(C)$ by their $\dep(v)=\min\{l_1,l_2,l_3\}$.
\smallskip

\noindent
\textbf{Step 1e}. 
The subtree $\core(C)\subset\MST(C)$ is formed by all the paths  $P_i$ and $P_{ij}$ from Definition~\ref{dfn:core} that have lengths more than $\be l(C)$, where $\be$ is a given branching factor.

\section{$\sk(C)$: Approximate Skeleton of a cloud $C\subset\R^m$}
\label{sec:ask}

The tree $\core(C)$ from Definition~\ref{dfn:core} has only few non-trivial vertices, but contains noisy zigzags with too many {\em trivial} vertices of degree~2.
This section discusses how to straighten these zigzags and decrease the total number of vertices.
\smallskip

We have tried Douglas-Peucker's heuristic \cite{DP73}, which was rather unstable and produced large zigzags on curved micelles in Fig.~\ref{fig:cylindrical+branched}.
The worst complexity is $O(n^2)$ in the number $n$ of points for $d>2$.
A final approximation can have a size $\Omega(n)$ even in $\R^2$.
Another problem with \cite{DP73} are potential self-intersections even in $\R^2$, which are caused by large zigzags that approximate non-monotone curves \cite{WSM04}.
\smallskip

The problem of straightening polygonal paths in a tree $\core(C)$ is harder than the curve simplification, because the input is a cloud of unorganized points.
So a final approximation should take into account the points of a cloud $C$ outside $\core(C)$. 
\medskip

\begin{dfn}
\label{dfn:monotone}
Let $L\subset\R^m$ be a straight line.
An ordered cloud $C=\{p_1,\dots,p_n\}\subset\R^m$ is called \emph{monotone} with respect to $L$ if the order of points is preserved by the orthogonal projection of $C$ to $L$.
\end{dfn}

Since there are many paths of $\core(C)$ to straighten, we split the cloud $C$ into monotone subclouds as formalized in Algorithm~\ref{alg:subclouds} in Appendix~A. 
Since monotone subpaths can be quickly found only in $\R^2$ \cite{shin2004optimal}, Theorem~\ref{thm:size} below will assume that each subpath $P$ between non-trivial vertices of $\core(C)$ is monotone by Definition~\ref{dfn:monotone} with respect to the straight line connecting the endpoints of $P$.
\smallskip

All results in this section are proved in Appendix A. 
Here are the Stage~2 steps.
\medskip

\noindent
\textbf{Step 2a}. 
Split every polygonal path between non-trivial vertices (of degrees $k\neq 2$) in the subtree $\core(C)\subset\MST(C)$ into monotone subpaths by Algorithm~\ref{alg:subclouds}.
\smallskip

\noindent
\textbf{Step 2b}. 
Each monotone subpath of $\core(C)$ with endpoints (say) $p_1,p_n$ has the subcloud $C'$ approximated by a polygonal path via points of $C'$ by Steps 2c--2f.
\smallskip

\noindent
\textbf{Step 2c}.
For each subcloud $C'=\langle p_1,\dots,p_n\rangle$ of points ordered by their orthogonal projections to $[p_1,p_n]$, start from $\ind(1)=1$ and find the next index $\ind(i)$ for $i=2,\dots,m$ by repeating Steps 2d--2e, which is  possible by Lemma~\ref{lem:algorithm} in Appendix~A.
\smallskip

\noindent
\textbf{Step 2d} (exponential). 
Find the smallest index $j$ such that $d( [p_{\ind(i-1)} p_l], C' )>\ep$ for $l=\ind(i-1)+2^{j+1}$, $j=0,1,2\dots$
For every index $l$, compute the distance $d( [p_{\ind(i-1)} p_l], C' )$ orthogonally to the line segment $[p_1 p_n]$ as in Definition~\ref{dfn:distances}.
\smallskip

\noindent
\textbf{Step 2e} (binary). 
Search for the maximum $\ind(i)$ between $\ind(i-1)+2^j$ and $\ind(i-1)+2^{j+1}$ such that $d( [p_{\ind(i-1)} p_{\ind(i)}], C' )\leq\ep$ by dividing the range in 2 halves.
\smallskip

\noindent
\textbf{Step 2f}. 
The found indices $\ind(i)$ specify a polygonal path $\ep$-approximating each monotone subcloud from Step~2b.
Combine all these paths into a full skeleton.
\smallskip

\noindent
\textbf{Step 2g}. 
Any edges of a length more than $\be l(C)$ from Definition~\ref{dfn:depth} are temporarily removed from the skeleton.
Each remaining connected component with only short edges is collapsed to its center of mass.
The resulting vertices are connected according to the temporarily removed edges to get the Approximate Skeleton $\sk(C)$.
\medskip

For a cloud $C\subset\R^m$, mark the endpoints of all monotone subpaths in $\core(C)$ obtained by Algorithm~\ref{alg:subclouds}.
Consider all skeletons $S\subset\R^m$ that have fixed vertices at the marked points of $C$ such that any polygonal path between fixed vertices (say $u,v$) is monotone under the orthogonal projection to the line segment $[u,v]$.
\smallskip

\noindent
\textbf{The approximation problem} for an error $\ep>0$ is to minimize the total number of vertices in a straight-line graph $S\subset\R^m$ whose each monotone path should $\ep$-approximate the corresponding subcloud of $C$ by the distance in Definition~\ref{dfn:distances}. 

\begin{thm}
\label{thm:size}
Let $k$ be the minimum number of vertices over all graphs $\ep$-approximating a given cloud $C\subset\R^m$. 
Then $\sk(C)$ lies in $C^{2\ep}$ and has at most $k$ vertices. 
\end{thm}

Theorem~\ref{thm:size} estimates the number of vertices of $\sk(C)$ when the geometric error is $2\ep$.
In practice, the tree $\core(C)$ has an initial approximation error for a given cloud $C$, because many points of $C$ may not be vertices of $\core(C)\subset\MST(C)$.
\smallskip

We measure the initial error $d(\core(C),C)$ by Definition~\ref{dfn:distances} and take the maximum of $d([v_iv_j],C)$ over monotone paths of $\core(C)$ computed in Algorithm~\ref{alg:subclouds}.
Stage~2 approximates $C$ by a graph simpler than $\core(C)$, but keeps the approximation error small.
The error $\ep$ in Corollary~\ref{cor:time} is $\ga\times d(\core(C),C)$, where $\ga$ is an error factor that takes values in the interval $[1.1,1.5]$ for the experiments in section~\ref{sec:experiments}.

\begin{cor}
\label{cor:time}
For any $n$ points $C\subset\R^m$ and any error factor $\ga>1$, an Approximate Skeleton $\sk(C)\subset\R^m$ within the $\ga d(\core(C),C)$-offset of the cloud $C$ (as in Definition~\ref{dfn:graph}) can be computed in time $O(\max\{c^6,c_p^2c^2_l) \}c^{10}n\log n\,\al(n))$, where $\al(n)$ is the inverse Ackermann function, 
 the constants $c,c_p,c_l$ are defined in Appendix~A.
\end{cor}

\section{Comparisons of 5 algorithms on real and synthetic data}
\label{sec:experiments}

This section experimentally compares the Approximate Skeleton $\sk(C)$ with those four skeletonization algorithms from section~\ref{sec:review} that have theoretical guarantees and accept any cloud $C$ of points: Mapper \cite{SMG07}, Metric Graph Reconstruction MGR \cite{ACCGGM12}, $\al$-Reeb graphs \cite{CHS15} and most recent discrete Morse theory (DMT) algorithm \cite{dey2018graph}.
\smallskip

The Mapper \cite{SMG07} is very flexible in the sense that its parameters might be manually tuned for given data over numerous clustering algorithms.
Having tried several possibilities, we have settled on the following choices from the original work \cite{SMG07}.
\smallskip

\noindent
1) Convert a cloud $C$ into a connected neighborhood graph $N(C)$ with Euclidean edge-lengths by using a distance threshold.
The filter function is the distance function in $N(C)$ from a root that is the furthest point from a random point in $C$.
\smallskip

\noindent
2) The image of the filter function is covered by 10 intervals with the 50\% overlap so that $C$ splits into 10 subclouds when filter values are in one of the 10 intervals.
\smallskip

\noindent
3) Each subcloud $C'$ is clustered by the single-linkage clustering with the threshold $\tau\times$the average edge-length of $\MST(C')$, where values of the factor $\tau$ are given in Table~\ref{tab:micelles}.
The final Mapper graph has a single node representing each cluster.
\medskip

The authors of the DMT algorithm by T.~Dey et al. \cite{dey2018graph} have kindly made their code available at https://github.com/wangjiayuan007/graph\_recon\_DM.
Starting from an unorganized cloud of points, e.g. centers of molecules of a micelle, we generated scalar values at nodes of a regular grid required for the DMT algorithm.
\smallskip

\noindent
1) We subdivide the axis-aligned bounding box of a cloud $C\subset\R^3$ into small boxes: minimum 20 rectangular boxes (as close to cubic as possible) along each side.
\smallskip

\noindent
2) The scalar values are found from the Kernel Density Estimate $KDE(p)=\sum\limits_{q\in C}\exp(-d(p,q))$ at every grid node $p$. 
The computed values are passed to the DMT with a parameter $\de$ that regulates how small density values are replaced by 0.
\smallskip

The MGR algorithm has required much more efforts, because the original code was lost as confirmed by the main author of \cite{ACCGGM12}. 
Since the algorithm was well-explained, we have implemented MGR ourselves and confirmed the earlier claim that ``it is often hard to find suitable parameters'' \cite[page~3]{GSBW11}.
Trying many values of the key parameter $r$ gave the zero success rate on the homeomorphism type. 
\smallskip 

Hence we have improved MGR by splitting this parameter into two: the first $r_1=15$ (values used in all experiments) was used for detecting vertex points, the second $r_2$ (three values $1,1.5,2$ in Table~\ref{tab:micelles} experiments) was used for clustering points of different types.
Only after using these different values, we have managed to push the success rates of MGR closer to $50\%$ on the homeomorphism type.
\smallskip 

The $\al$-Reeb graph has the essential parameter $\al$ whose values $20,25,30$ were tried in all experiments.
$\sk(C)$ has little dependence on the branching factor $\be$, e.g. all values $[20,50]$ produced almost identical results in Table~\ref{tab:micelles} and Fig.~\ref{fig:synthetic}.
\smallskip

Since three output graphs (Mapper, $\al$-Reeb and MGR) are abstract, to compute any geometric error of approximation, we map them to $\R^3$ by sending every node $v$ of $G$ to the average point (center of mass) of the cluster (for Mapper and MGR) or subgraph (for $\al$-Reeb) corresponding to $v$. 
Each link between nodes is mapped as a line segment between the corresponding points in $\R^3$.
\smallskip

\begin{figure*}[t]
\includegraphics[width=0.33\textwidth]{cylindrical/cylindrical_cloud_blue.png}
\includegraphics[width=0.33\textwidth]{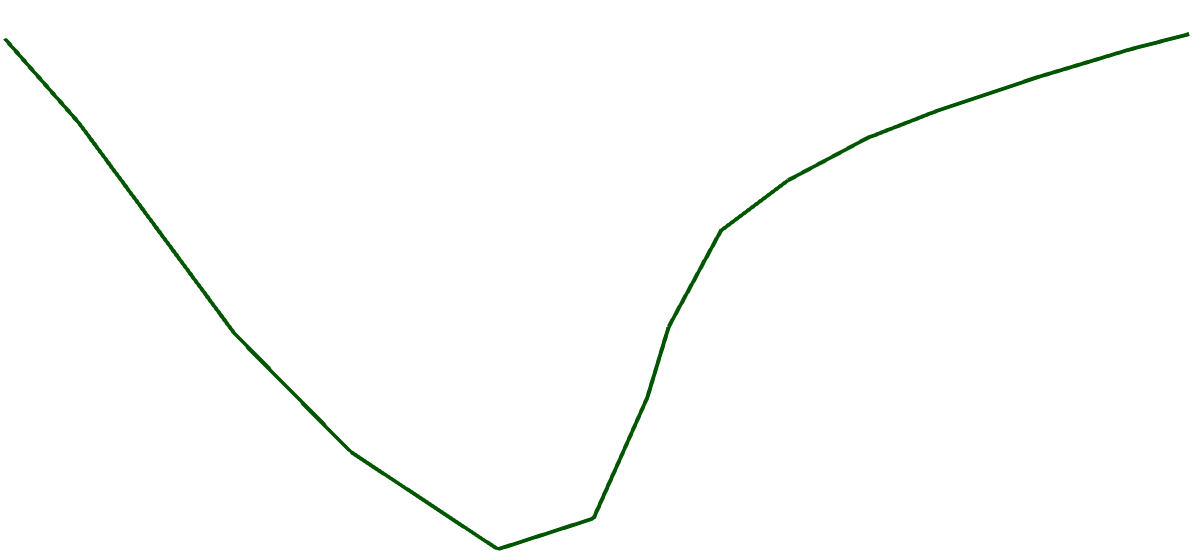}
\includegraphics[width=0.33\textwidth]{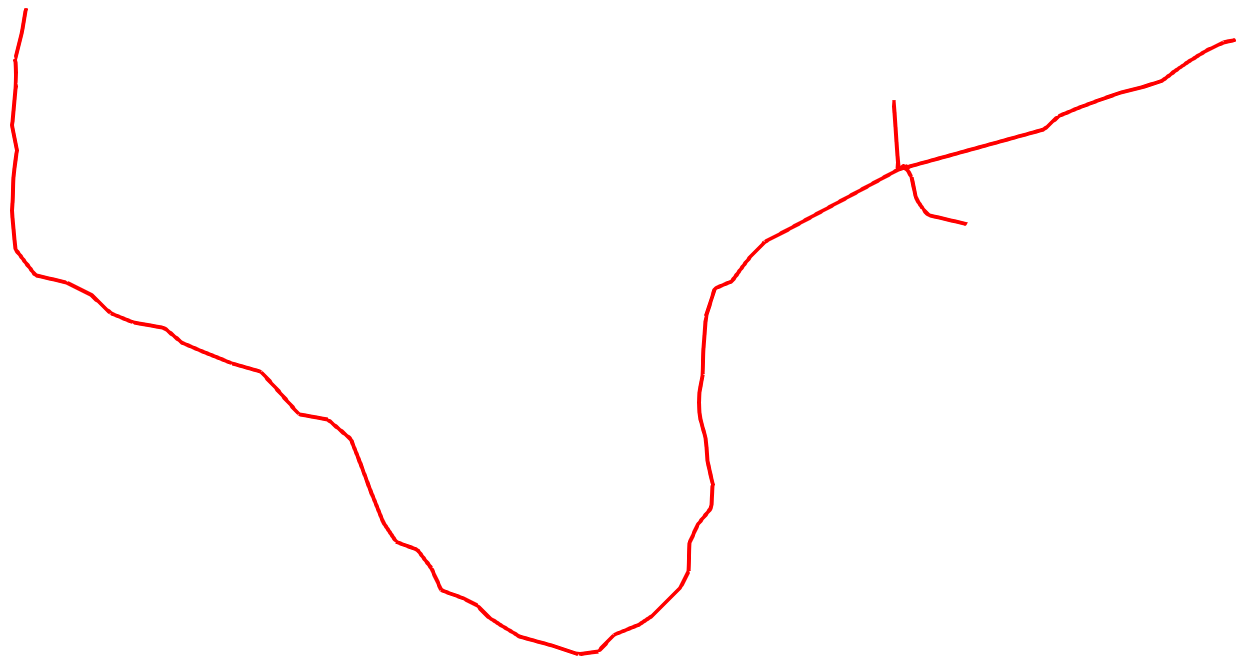}

\includegraphics[width=0.33\textwidth]{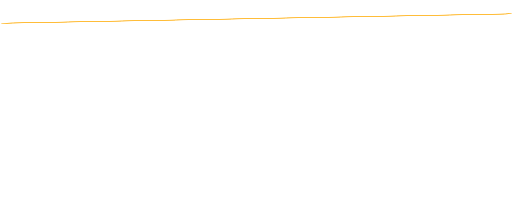}
\includegraphics[width=0.33\textwidth]{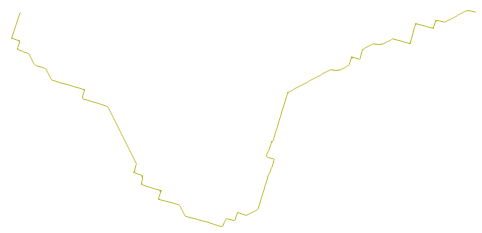}
\includegraphics[width=0.33\textwidth]{cylindrical/cylindrical_ASk_blue.png}
\caption{\textbf{1st}: a cylindrical micelle with no branching vertices. 
\textbf{2nd}: Mapper, 
\textbf{3rd}: $\al$-Reeb,
\textbf{4th}: MGR, 
\textbf{5th}: DMT, 
\textbf{6th}: new $\sk(C)$.
}
\label{fig:cylindrical_outputs}
\end{figure*}

\begin{figure*}[h!]
\includegraphics[width=0.33\textwidth]{branched/branched_cloud_blue.png}
\includegraphics[width=0.33\textwidth]{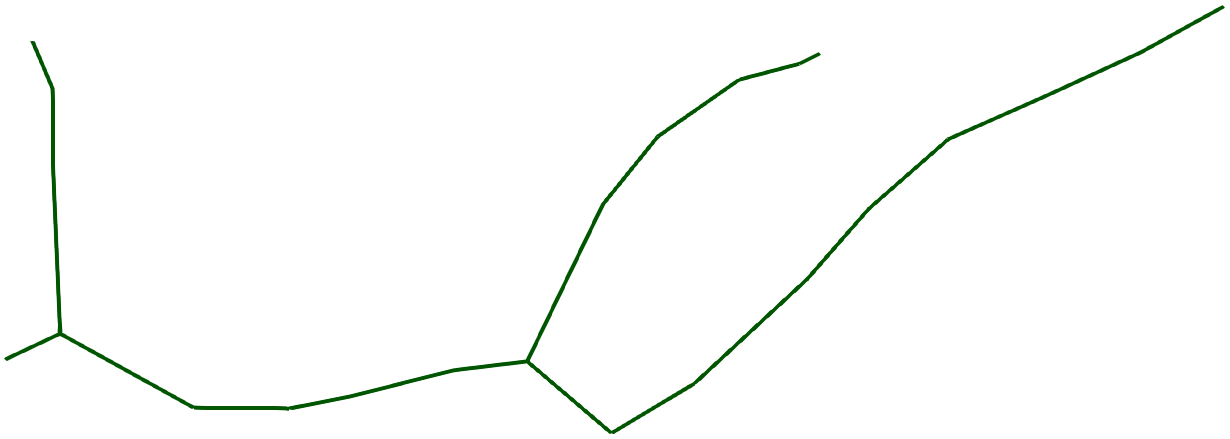}
\includegraphics[width=0.33\textwidth]{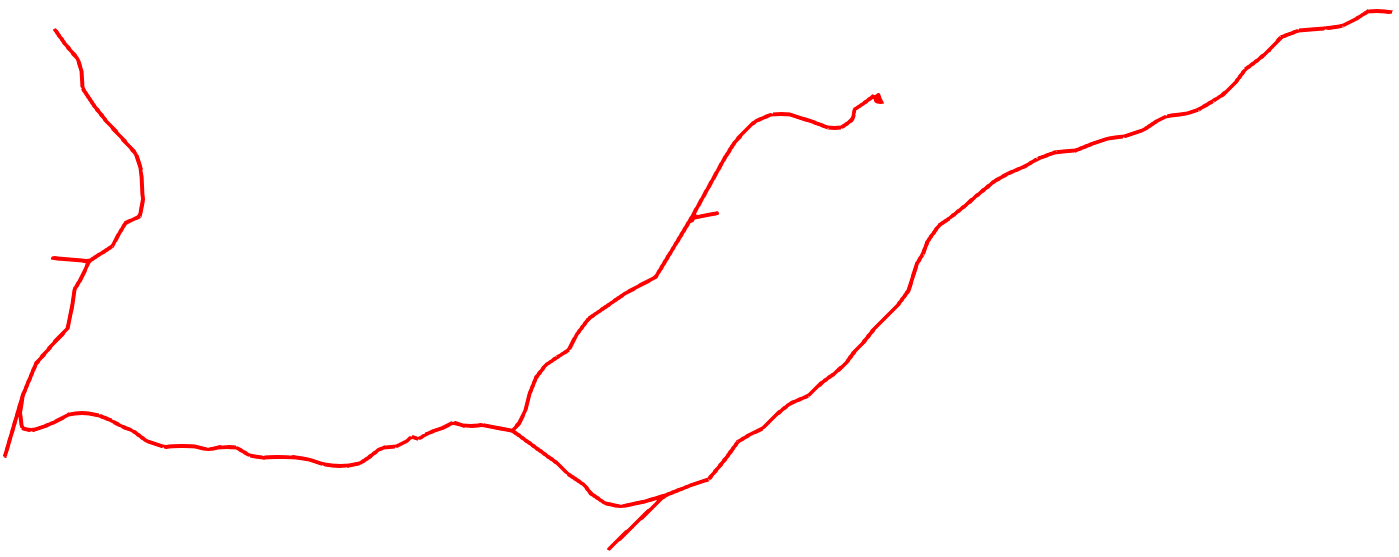}

\includegraphics[width=0.33\textwidth]{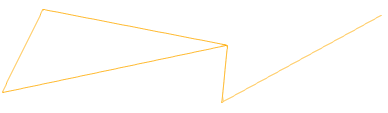}
\includegraphics[width=0.33\textwidth]{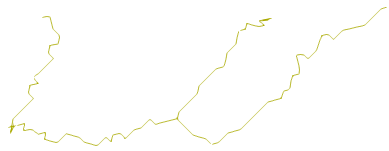}
\includegraphics[width=0.33\textwidth]{branched/branched_ASk_blue.png}
\caption{\textbf{1st}: a branched micelle with exactly one degree 3 vertex, 
\textbf{2nd}: Mapper, 
\textbf{3rd}: $\al$-Reeb,
\textbf{4th}: MGR, 
\textbf{5th}: DMT, 
\textbf{6th}: new $\sk(C)$.
}
\label{fig:branched_outputs}
\end{figure*}

\begin{figure*}[h!]
\includegraphics[width=0.33\textwidth]{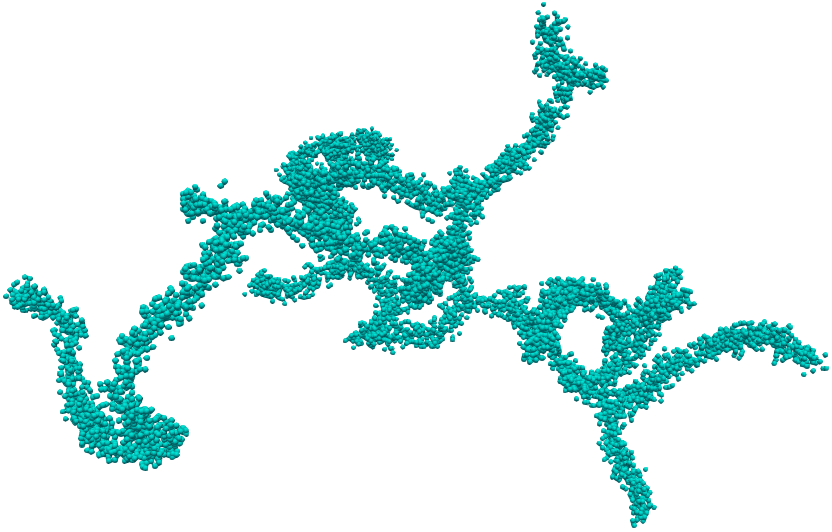}
\includegraphics[width=0.33\textwidth]{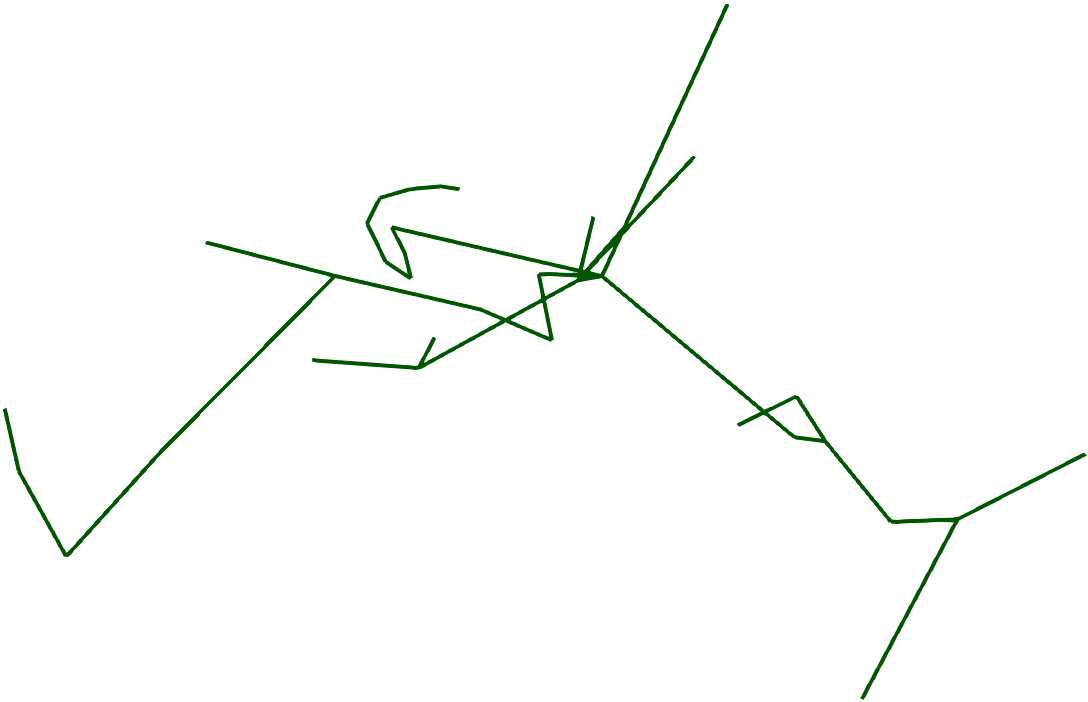}
\includegraphics[width=0.33\textwidth]{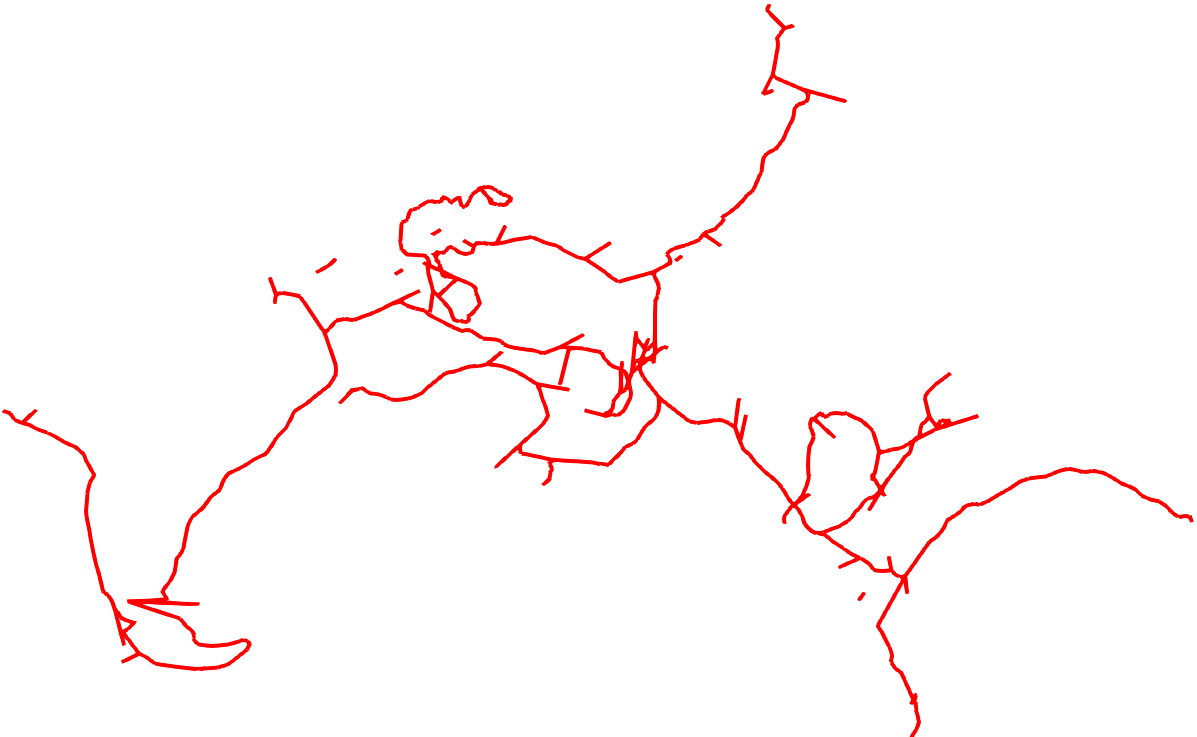}

\includegraphics[width=0.33\textwidth]{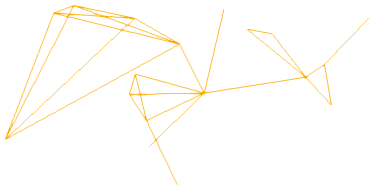}
\includegraphics[width=0.33\textwidth]{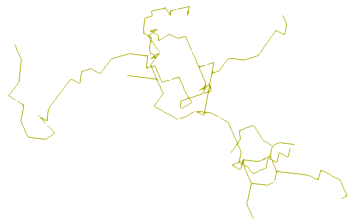}
\includegraphics[width=0.33\textwidth]{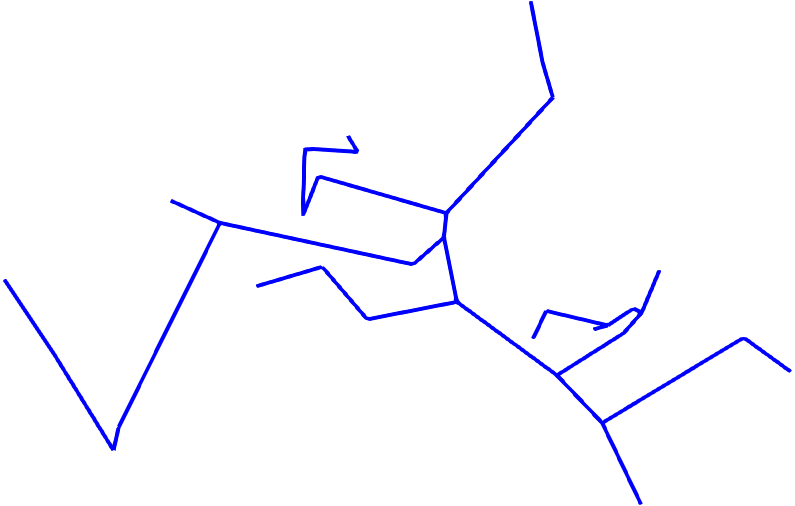}
\caption{\textbf{1st}: `Christmas tree' micelle with several degree 3 vertices). 
\textbf{2nd}: Mapper, 
\textbf{3rd}: $\al$-Reeb,
\textbf{4th}: MGR, 
\textbf{5th}: DMT, 
\textbf{6th}: new $\sk(C)$.
All intersections come only from planar projections.
}
\label{fig:Christmas_outputs}
\end{figure*}

\begin{figure*}[t]
\includegraphics[width=0.33\textwidth]{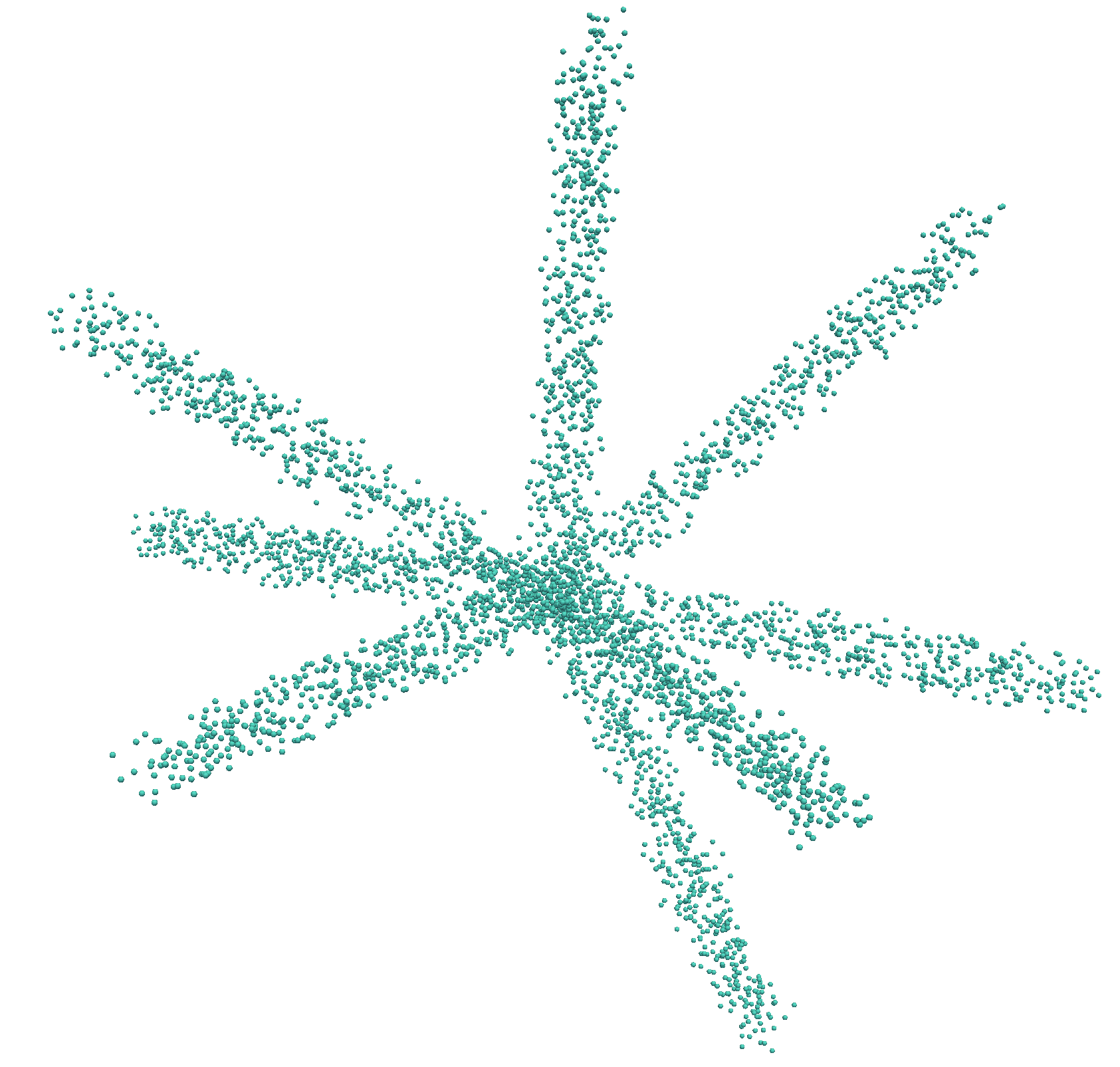}
\includegraphics[width=0.33\textwidth]{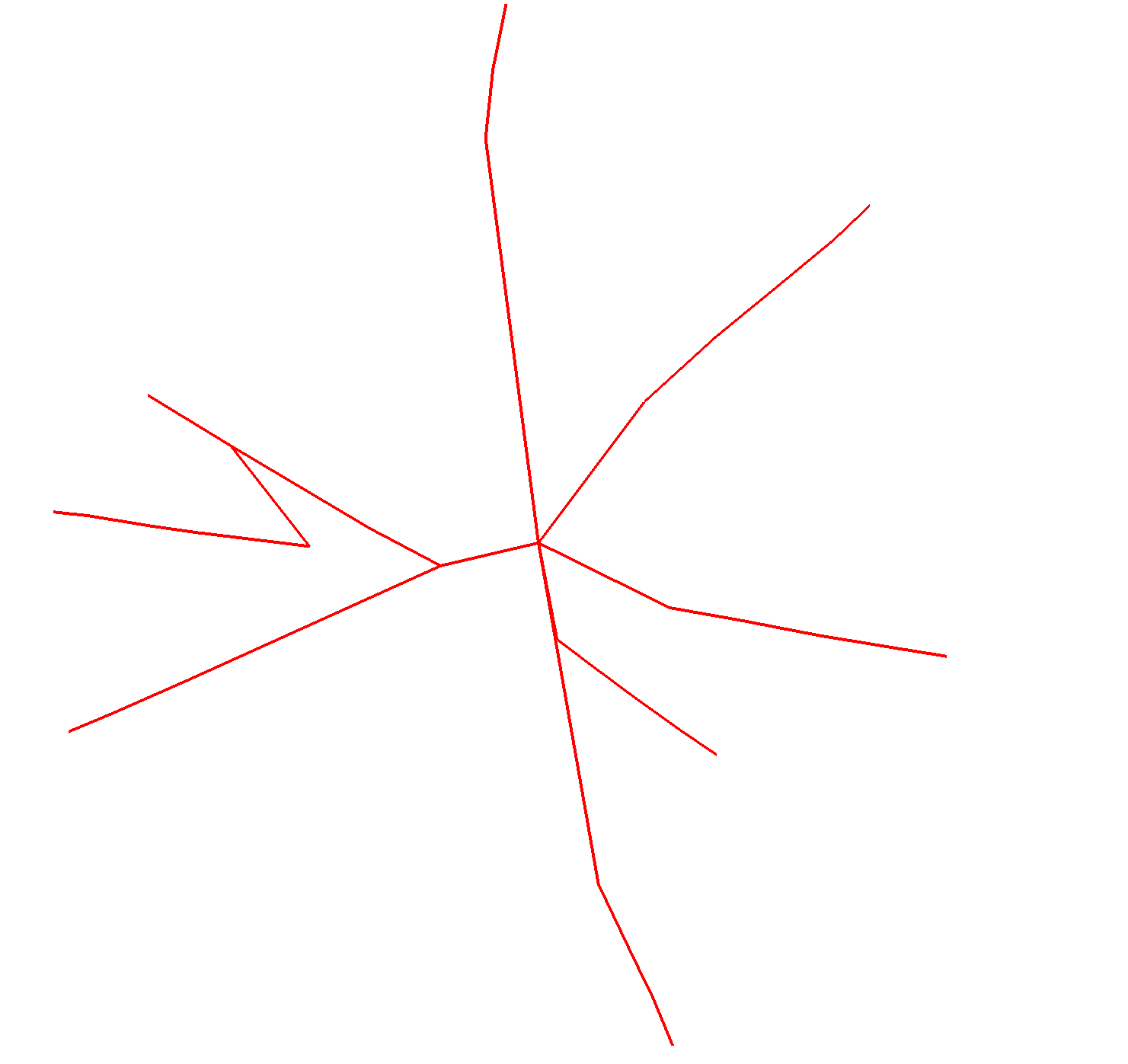}
\includegraphics[width=0.33\textwidth]{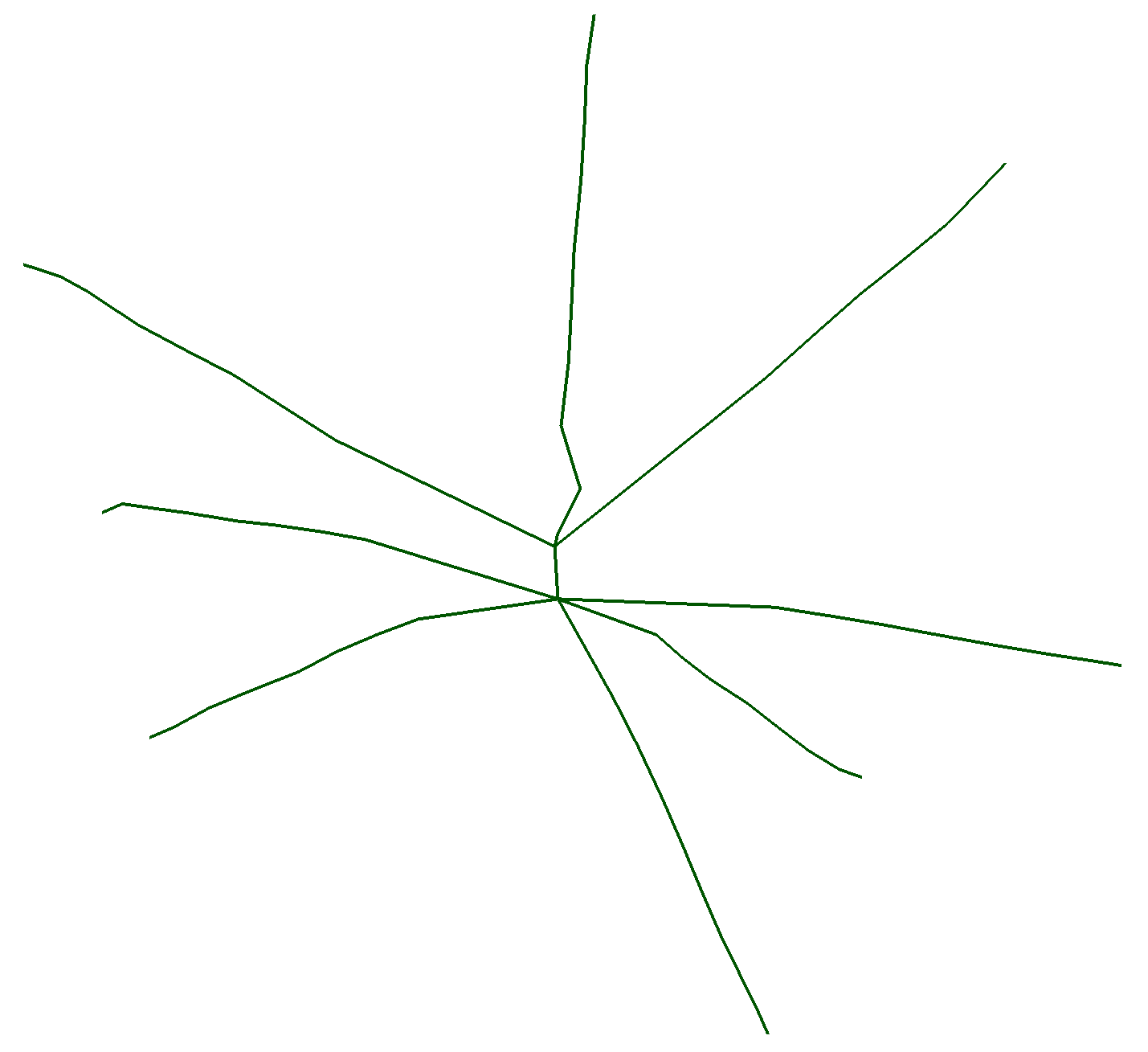}

\includegraphics[width=0.33\textwidth]{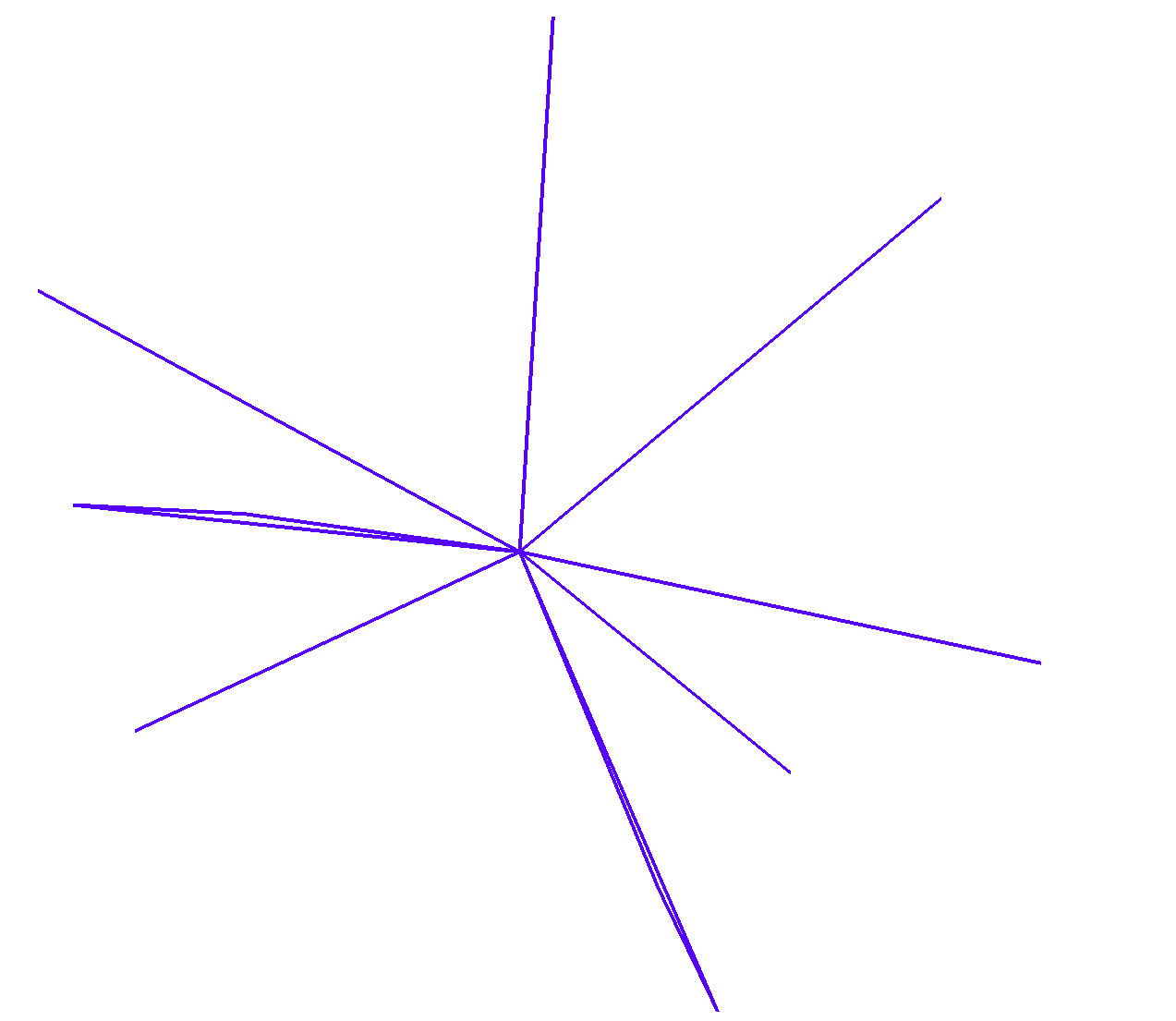}
\includegraphics[width=0.33\textwidth]{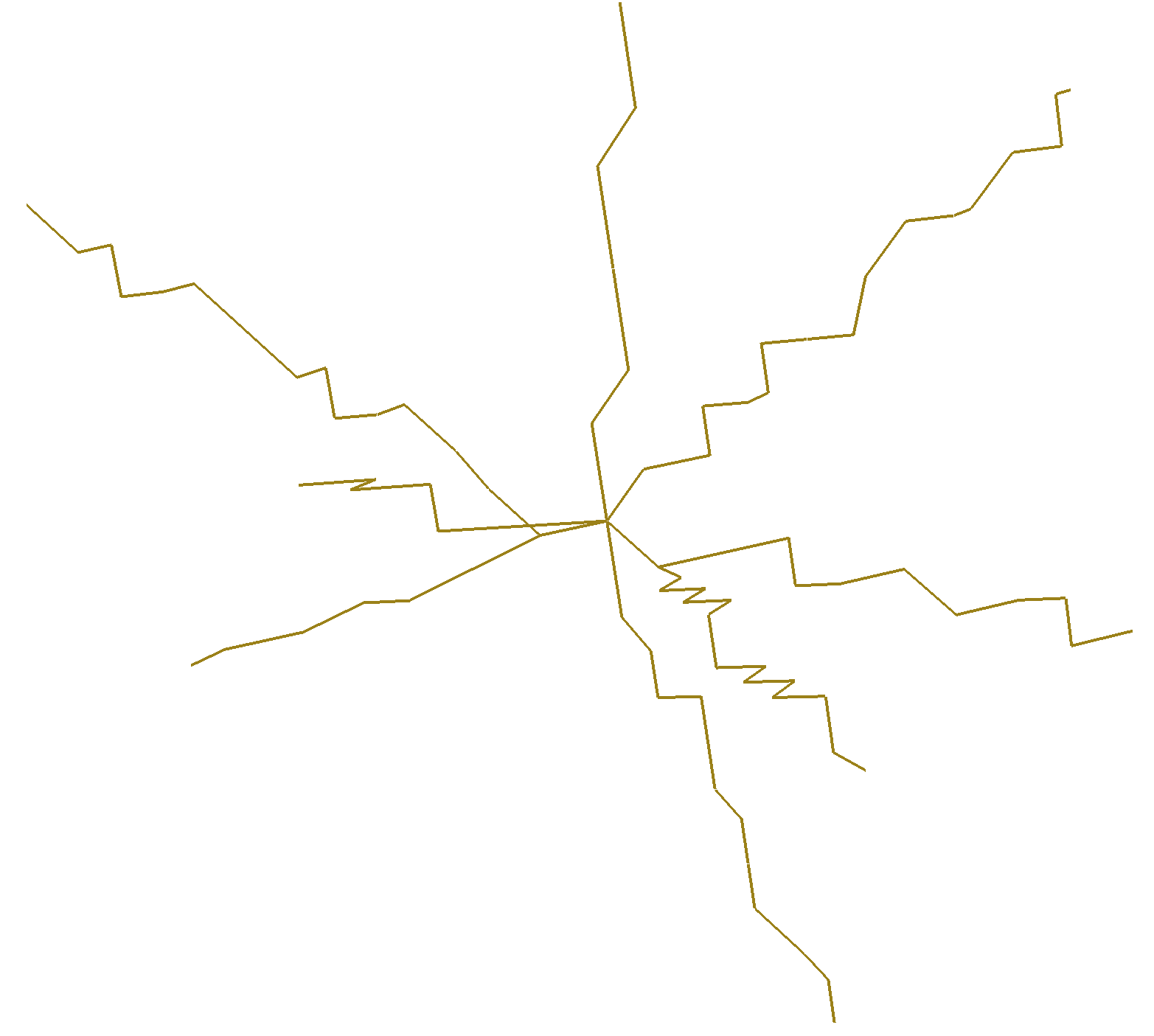}
\includegraphics[width=0.33\textwidth]{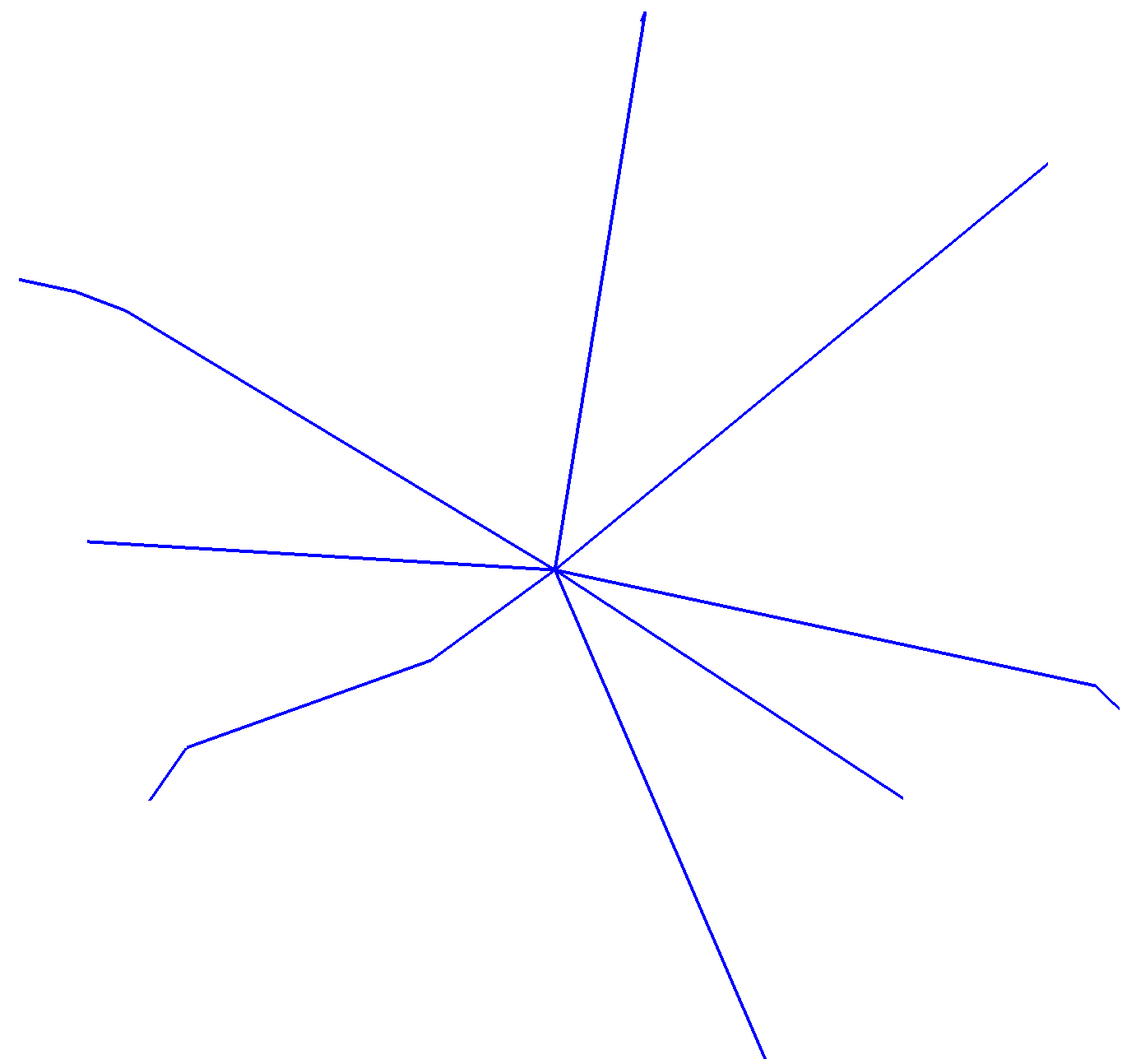}
\caption{\textbf{1st}: a random point sample around an 8-star in $\R^3$,
\textbf{2nd}: Mapper, 
\textbf{3rd}: $\al$-Reeb,
\textbf{4th}: MGR, 
\textbf{5th}: DMT, 
\textbf{6th}: new $\sk(C)$.
}
\label{fig:8star_outputs}
\end{figure*}

\begin{figure*}[t]
\includegraphics[width=0.48\textwidth]{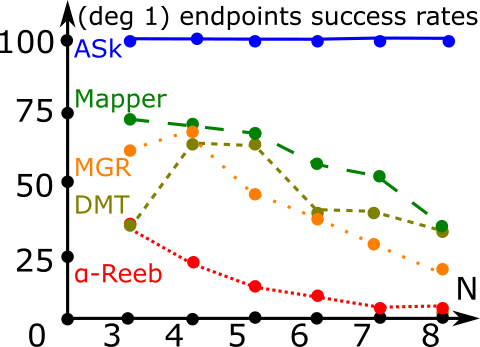}
\includegraphics[width=0.48\textwidth]{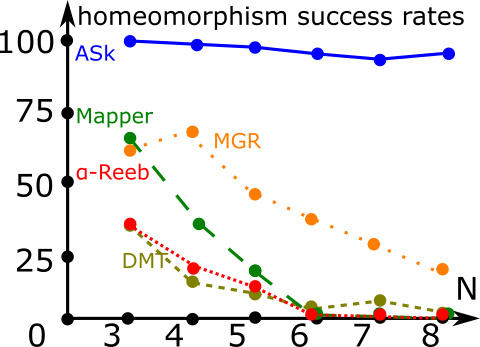}
\medskip

\includegraphics[width=0.48\textwidth]{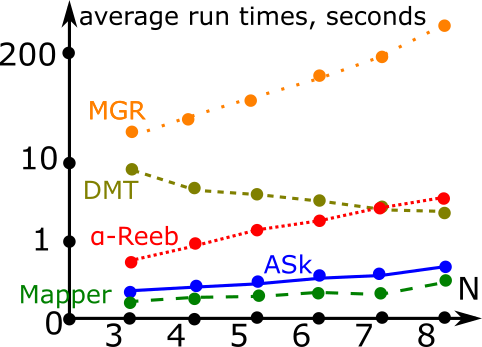}
\includegraphics[width=0.48\textwidth]{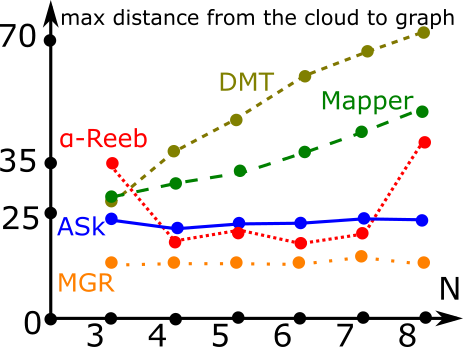}
\caption{For $N=3,\dots,8$, each dot represents the average over 100 noisy samples around a random $N$-star graph in $\R^3$ and 3 parameters as in Table~\ref{tab:micelles}. 
Mapper: green and long-dashed, MGR: orange and sparsely dotted, $\al$-Reeb: red and densely dotted, DMT: olive and short-dashed, $\sk(C)$: blue and solid.
\textbf{Top left}: the success rate in percentages for detecting a correct number of endpoints. 
\textbf{Top right}: the success rate in percentages for detecting the homeomorphism type. 
\textbf{Bottom left} (logarithmic scale): average run times in milliseconds.
\textbf{Bottom right}: the max distance from a cloud $C$ to reconstructed graphs. 
The exact numbers are in the txt files in the supplementary materials.
}
\label{fig:synthetic}
\end{figure*}

Figures~\ref{fig:cylindrical_outputs}--\ref{fig:8star_outputs} show clouds and outputs of 5 algorithms.
Since real micelles have irregular shapes in $\R^3$, their 2D projections may contain intersections of edges.
\smallskip

Table~\ref{tab:micelles} shows the average results of the three algorithms on the dataset of more than 100 real micelles (clouds of about 300 molecules) whose endpoints and homeomorphism types were manually detected.
A \emph{homeomorphism} is a 1-1 continuous map with a continuous inverse, so a homeomorphism type is a stronger shape descriptor than a homotopy type, which counts only linearly independent cycles.
\smallskip

\smallskip

The \emph{most important error} measure for the tree reconstruction problem in section~\ref{sec:intro} is the success rate for detecting a correct homeomorphism type.
Indeed, an incorrect graph can be perfect on other errors, e.g. $\MST(C)$ is extremely fast, has the zero geometric error (for many distances between a cloud and a reconstructed graph) and even has a correct homotopy type (no cycles) for any underlying tree $T$.
\smallskip

Hence the key results are in the middle column of Table~\ref{tab:micelles} and the top right picture of Fig.~\ref{fig:synthetic}.
We included the success rate on the number of endpoints (degree~1 vertices) as a weaker topological error.
As $\MST(C)$ shows above, only if an algorithm performs well on a topological reconstruction, it makes sense to evaluate the performance on other measures such as geometric distances and time.  
\smallskip

\begin{table}[t]
\caption{ 
Columns 3-4 contain success rates for detecting the correct number of endpoints and a homeomorphism type of a graph over more than 100 real micelles.
Column 5 contains the maximum Euclidean distance from points of a given cloud $C$ to the reconstructed graph $G\subset\R^3$.}
\label{tab:micelles}
\begin{tabular}{c|c|c|c|c|c}
\toprule
algorithm & parameters & endpoints success & homeomorphism success & time, ms  & max distance from $C$
\\
\midrule
Mapper & $\tau=1.25$ & 54.21\% & 54.21\% & \textbf{18}  & 4.59 \\
Mapper & $\tau=1.75$ & 66.36\% & 66.36\%  & \textbf{18} & 4.62 \\
Mapper & $\tau=2.25$ & 68.20\% & 68.20\%  & 19 & 4.67 \\
\midrule
MGR & $r_2=1$ & 48.60\% & 45.79\%  & 25010 & 5.95 \\
MGR & $r_2=1.5$ & 40.19\% & 40.19\% & 17410 & 5.51 \\
MGR & $r_2=2$ & 29.91\% & 29.91\% & 25480 & \textbf{3.46} \\
\midrule
$\al$-Reeb & $\al=20$ & \textbf{98.13\%} & \textbf{98.13\%}  & 367 & 10.49 \\
$\al$-Reeb & $\al=25$ & 97.20\% & 97.20\%  & 375 & 12.50 \\
$\al$-Reeb & $\al=30$ & \textbf{98.13\%} & \textbf{98.13\%}  & 373 & 14.19 \\
\midrule
DMT & $\de=0.1$ & 48.60\% & 45.79\% & 6290 & 5.95 \\
DMT & $\de=0.2$ & 40.19\% & 40.19\% & 6192 & 5.51 \\
DMT & $\de=0.3$ & 29.91\% & 29.91\% & 6410 & \textbf{3.46} \\
\midrule
$\sk(C)$ & $\be=20$ & \textbf{98.13\%} & \textbf{98.13\%} & 42 & 5.16 \\
$\sk(C)$ & $\be=30$ & \textbf{98.13\%} & \textbf{98.13\%} & 42 & 5.16 \\
$\sk(C)$ & $\be=40$ & 97.20\% & 97.20\% & 42 & 5.31 \\
\bottomrule
\end{tabular}
\end{table}

Table~\ref{tab:micelles} shows that the Mapper, MGR and DMT essentially depend on their parameters, because the success rates, run time and distance error significantly vary when the parameters are only slightly changed.
The $\al$-Reeb and $\sk$ were stable, because Table~\ref{tab:micelles} contains almost identical success rates for different parameters.
\smallskip

Both algorithms achieved best results on the most important measure of the homeomorphism success rate, though the top right picture in Fig.~\ref{fig:synthetic} highlights $\sk(C)$ and MGR as the best for homeomorphism.
In comparison with $\al$-Reeb and MGR, the Approximate Skeleton $\sk(C)$ is much faster and achieves similar distance errors, see the relevant results in both Table~\ref{tab:micelles} and Fig.~\ref{fig:synthetic}.
\smallskip

In addition to the comparison on more than 100 micelles, we have tested the algorithms on the much larger dataset of synthetic clouds generated as follows.
\smallskip

\noindent
1) An \emph{$N$-star} in $\R^3$ has one vertex at $0\in\R^3$ and straight edges of length 100 to $N$ {\em endpoints}in random directions with a minimum angle $\dfrac{\pi}{4}$ between edges.
\smallskip

\noindent
2) For $N=3,\dots,8$ and every of 100 random $N$-stars $T$, we found a minimum axis-aligned box containing $T$, enlarged this box by the noise bound of 10\%.
\smallskip

\noindent
3) We uniformly chose a random point $p$ in the resulting box and checked if $p$ is at a distance at most 10 (=10\% of edge-lengths) from $T$.
If successful, $500N$ such points form a noisy sample of the ground truth $N$-star $T\subset\R^3$.
\smallskip

Fig.~\ref{fig:synthetic} shows 4 plots for the 4 error measures of 5 algorithms, which were averaged over 3 values of essential parameters as in Table~\ref{tab:micelles}. 
The Mapper threshold factor for single-edge clustering was $\tau\in\{1.25,1.75,2.25\}$.
The $\al$-Reeb scale was $\al\in\{20,25,30\}$.
The branching factor of $\sk(C)$ was $\be\in\{20,30,40\}$.
\smallskip

For the correct number of endpoints, the new skeleton $\sk(C)$ achieves 100\% results on the synthetic clouds, because Definition~\ref{dfn:depth} provides a very stable concept of a deep vertex not critically depending on a branching factor $\be$.
For the homeomorphism type, the minimum success of $\sk(C)$ is 96\%, because all short branches of $\MST(C)$ are removed to get $\core(C)$ homeomorphic to an underlying tree.
\smallskip

For the random point sample of the 8-star graph in Fig.~\ref{fig:8star_outputs}, the 2nd, 3rd and 5th graphs have several branched vertices instead of one. 
The 5th graph has several zigzags, which would be straightened in $\core(C)$.
The 4th graph has a triangular cycle because of incorrectly detected overlaps of clusters corresponding to vertices.

\section{Conclusions and a discussion of the Approximate Skeleton}
\label{sec:conclusions}
 
Though the current implementation was tested in $\R^3$, all steps and results work in any $\R^m$.
Here is the summary of the key contributions to data skeletonization.
\smallskip

\noindent
$\bullet$
The detection of deep (branched) vertices in Definition~\ref{dfn:depth} uses a global structure of longest paths within $\MST(C)$, hence is more stable under a change of parameters.
\medskip

\noindent
$\bullet$
To improve the Metric Graph Reconstruction by M.~Aanjaneya et al. \cite{ACCGGM12}, we have split one parameter $r$ (used for detecting vertex points and also for clustering later) into two separate parameters (with default values) $r_1=15$, $r_2\in[1,2]$, which led to more successful (20-40\% rates instead of 0\%) reconstructions in Table~\ref{tab:micelles}.
\medskip

\noindent
$\bullet$
Theorem~\ref{thm:size} proves the first size guarantees (on a small number of vertices) for the Approximate Skeleton $\sk(C)$, while all past methods from section~\ref{sec:review} considered topological (mostly homotopy type) or metric properties of reconstructed graphs.
\medskip

\noindent
$\bullet$
Corollary~\ref{cor:time} says that the Approximate Skeleton $\sk(C)$ can be quickly computed within a given error as required in the Tree Reconstruction Problem from section~\ref{sec:intro}.
\medskip

Because of the page limit the last author couldn't include one more result on $\sk(C)$ with realistic conditions on an underlying tree $T\subset\R^m$ and its noisy sample $C$ to guarantee that $\MST(C)$ and $\sk(C)$ are homeomorphic to $T$.
This is the first advance after Giesen's guarantees for shortest paths through sample points \cite{giesen1999curve} in 1999.
The C++ code of $\sk(C)$ is at https://github.com/YuryUoL/AsKAlgorithm.
\smallskip

In comparison with the past methods in section~\ref{sec:review}, $\sk(C)$ starts from the most challenging input (an unorganized cloud of points $C\subset\R^m$ without any extra structure such a metric graph or a regular grid or a mesh), outputs an embedded graph in $\R^m$ and provides two guarantees: combinatorial in Theorem~\ref{thm:size} and geometric in Corollary~\ref{cor:time} and topological.
Appendix~A has all missed proofs.
Appendix~B includes more experiments.
We thank all reviewers for their helpful suggestions.

\bibliographystyle{spmpsci} 
\bibliography{approximate_skeleton}       

\section{Appendix A: proofs of all the statements from section~\ref{sec:ask} }

The proof of Corollary~\ref{cor:time} below uses Algorithm~\ref{alg:depths} for depths of vertices in $\MST(C)$ from Definition~\ref{dfn:depth}.
The depth is trivial (equal to 0) for any degree~1 vertex.
For any other vertex $v$, the depth can be recursively computed from lengths of edges at $v$ and depths of neighbors of $v$.
Imagine a water flow simultaneously starting from all degree~1 vertices of $\MST(C)$ and moving towards internal vertices inside $\MST(C)$.
At every vertex $v$ of degree $k\geq 3$, the flow waits until $v$ is reached from $k-1$ directions (edges at $v$), then the flow moves further in the remaining $k$-th direction.
\medskip

\begin{algorithm}[b]
   \caption{Computing depths of vertices from Definition~\ref{dfn:depth} in Step~1c by `simultaneous flows' moving from endpoints.}
   \label{alg:depths}
\begin{algorithmic}
   \STATE \textbf{Input:} the initial tree $T=\MST(C)$
   \STATE {Initialize Minimal Binary Heap $H$ of (vertex,depth)}
   \STATE{For all deg~1 vertices $v\in T$, add $(v,0)$ to $H$};
   \WHILE {H is not empty}
   \STATE $(v,d)$ = H.pop(); // take the vertex $v$ of a min depth
   \STATE set $u$ = the only neighbor of $v$ in $T$;
   \STATE $d_{new} = d + \mbox{edge-length of }(u,v)$;
   \STATE Remove the edge $uv$ from $T$, but keep $u,v\in T$;
   \STATE Add $(v,d_{new})$ to front of the list Neighbors$(u)$;
   \IF{$\text{deg}(u) = 0$ in $T$}
   \STATE Set $flag[u] = true$;
   \ELSIF{$\text{deg}(v) = 1 $}
   \STATE Add the pair $(u,d_{new})$ to the heap $H$
   \ENDIF
   \ENDWHILE
   \STATE Initialize a vector[] depths; // a future output
   \FOR{ all $\text{deg}(v) > 2$ vertices $v\in\MST(C)$}
   \IF {$flag[v] = true$}
   \STATE Set depths[v] = 3rd element of Neighbors$(v)$;
   \ELSE
   \STATE Set depths[v] = 2nd element of Neighbors$(v)$;
   \ENDIF 
   \ENDFOR
\end{algorithmic}
\end{algorithm}
\medskip

Definition~\ref{dfn:distances} introduces a new distance between a cloud $C\subset\R^m$ and a polygonal line.
Recall that $d$ is the Euclidean distance in $\R^m$.
We assume that the points $C=\langle p_1,\dots,p_n\rangle\subset\R^m$ are ordered by their orthogonal projections to the line $[p_1 p_n]$.

\begin{figure}[h]
\includegraphics[width=\linewidth]{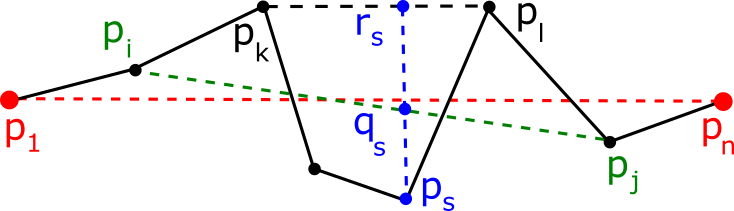}
\caption{The distance from $p_m$ to $[p_ip_j]$ in Definition~\ref{dfn:distances} is measured orthogonally to $[p_1 p_n]$.}
\label{fig:segment_approximation}
\end{figure}

\begin{dfn}
\label{dfn:distances}
For $1\leq i<s<j\leq n$, let $H(p_s)\subset\R^m$ be the hyperspace that is orthogonal to $[p_1p_n]$ and passes through $p_s$.
The {\em distance} between $p_s$ and $[p_i p_j]$ is measured orthogonally to $[p_1 p_n]$ as $d(p_s,[p_i p_j]) = d( p_s, H(p_s)\cap [p_i p_j] )$, see Fig.~\ref{fig:segment_approximation}.
Consider the {\em distance} $d([p_i p_j],C) = \max\limits_{i<s<j} d( p_s, H(p_s)\cap [p_i p_j] )$.
For $1\leq\ind(1)<\dots<\ind(k)\leq n$, the {\em distance} between $C$ and the polygonal line $P=\langle p_{\ind(1)},\dots, p_{\ind(k)}\rangle$ is defined as $d(P,C)=\max\limits_{2\leq i\leq k} d([p_{\ind(i-1),\ind(i)}], C)$.
\end{dfn}

Lemma~\ref{lem:algorithm} below justifies the steps of Stage~2 in section 4, which outputs the Approximate Skeleton $\sk(C)$ starting from $\core(C)$ obtained in Stage~1 at the end of section~\ref{sec:core}.
\smallskip

\begin{lem}
\label{lem:algorithm}
Let $C=\langle p_1,\dots,p_n\rangle$ be points ordered according to their orthogonal projections to $[p_1,p_n]$.
For $\ep>0$, one can find indices $1=\ind(1)<\dots<\ind(m)=n$ in time $O(n\log n)$ so that the estimates below for the distances in Definition~\ref{dfn:distances} hold:
\smallskip

\noindent
(a)
$d( [p_{\ind(i-1)} p_k], C )\leq\ep$ for $\ind(i-1)<k\leq\ind(i)$,
\smallskip

\noindent
(b)
$d( [p_{\ind(i-1)} p_{\ind(i)+1}], C )>\ep$ for any $1<i<m$ .
\end{lem}

\begin{algorithm}
\label{alg:subclouds}
\caption{ 
\emph{monotone} subclouds of $C$.
For each point $p$ in a given cloud $C\subset\R^m$, find approximately its closest edge of $\core(C)$.
Then any edge $e\subset\core(C)$ has the {\em edge-cloud} $C(e)\subset C$ of points that are closer to $e$ than to other edges of $\core(C)$.
For every polygonal path $v_1,\dots,v_k$ between non-trivial vertices $v_1,v_k\in\core(C)$ define cloud $Y = \cup_{i=1}^{k-1} C((v_i,v_{i+1}))$ and straight line $L$ spanned by points $v_1$ and $v_k$. Run the algorithm above with parameters $(Y,L)$.}

\begin{algorithmic}
   \STATE \textbf{Input:} unordered set of points $s$ and line $l$
   \STATE \textbf{Output:} Sequence of vertices that form monotone paths $k$ and $t$ containing ordering of points $s$.
   \STATE Define $t$ to be ordering of $s$ obtained from projecting $s$ to $l$ orthogonaly.
   \STATE Define $q$ to be the map from points $s$ to their indices in $t$.
   \STATE Define $k$ to be queue and add first point of $s$ to $k$.
   \STATE Define $a$ to be the first point of $s$.
   \WHILE{Point $a$ is not the last point of $t$ }
    \STATE Let point $b$ be the next point from $a$ in ordering $t$
   \WHILE{Point $b$ is not the last point of $t$ and $k[a] < k[b]$}
   \STATE Set $a$ to be $b$.
   \STATE Set $b$ to be the next point from $b$ in ordering $t$.
   \ENDWHILE
   \STATE Add $b$ to $k$.
   \IF {$b$ is the last point in ordering $t$}
   \STATE Exit the program.
   \ENDIF
   \WHILE{Point $b$ is not the last point of $t$ and $k[b] < k[a]$}
    \STATE Set $a$ to be $b$.
   \STATE Set $b$ to be the next point from $b$ in ordering $t$.
   \ENDWHILE
    \STATE Add $b$ to $k$.

   \ENDWHILE
\end{algorithmic}
\end{algorithm}

The following lemma is needed for Theorem~\ref{thm:size} and is conveniently illustrated in Fig.~\ref{fig:segment_approximation} below.
Recall that the distances from Definition~\ref{dfn:distances} are computed orthogonally to the straight segment $[p_1,p_n]$ passing through the endpoints of a monotone point cloud $C$.

\begin{lem}
\label{lem:segment_error}
Let $C=\langle p_1,\dots,p_n\rangle$ be points ordered according to their orthogonal projections to $[p_1 p_n]$.
Then $d([p_k p_l],C)\leq 2d([p_i p_j],C)$ for any indices $i\leq k< l\leq j$. 
\end{lem}
{\em Proof.}
For any $k<m<l$, let $H(p_m)\subset\R^m$ be the hyperspace that is orthogonal to $[p_1p_n]$ and passes through $p_m$.
Consider the intersection points $q_m = H(p_m)\cap [p_i p_j]$ and $r_m= H(p_m)\cap [p_k p_l]$. 
Let $\ep = d([p_i p_j],C)$, then $d(p_m, q_m)=d(p_m,[p_i p_j])\leq\ep$.
Since the points $p_k$ and $p_l$ are $\ep$-close to the segment $[p_i p_j]$, the
intermediate point $r_m\in[p_k p_l]$ is also $\ep$-close to $[p_i p_j]$, i.e. $d(r_m,q_m)=d(r_m,[p_i p_j])\leq\ep$.
The triangle inequality implies that $d(p_m,r_m)\leq d(p_m,q_m) + d(q_m,r_m)\leq 2\ep$.
Taking the maximum over $k<m<l$, we get $d([p_k p_l],C)\leq 2\ep$.
\qed
\bigskip

\noindent
{\em Proof of Lemma~\ref{lem:algorithm}.}
Assuming that indices $1=\ind(1)<\dots<\ind(i-1)$ were found, we search for the next index $\ind(i)$ as follows.
Search exponentially by trying indices $k=\ind(i-1)+2^j$ for $j=0,1,\dots$ while $d( [p_{\ind(i-1)} p_k], C )\leq\ep$. 
\smallskip

Each evaluation of the distance $d( [p_{\ind(i-1)} p_k], C )$ requires $O(k-\ind(i-1))$ time, because we need to compare $k-\ind(i-1)-1$ distances to 
$[p_{\ind(i-1)} p_k]$ (orthogonally to $[p_1 p_n]$) from every point of $C$ between $p_{\ind(i-1)}$ and $p_k$. 
\smallskip

After finding $k=\ind(i-1)+2^j$ and $l=\ind(i-1)+2^{j+1}$ such that $d( [p_{\ind(i-1)} p_k], C )\leq\ep$ and $d( [p_{\ind(i-1)} p_l], C )>\ep$, 
we start a binary search for $\ind(i)$ in the range $[k,l)$ each time choosing one half of the current range until both conditions (a)-(b) hold.
\smallskip

Finding the next index $\ind(i)$ requires $O(\log n)$ computations for the distance $d( [p_k p_l], C )$, where $l-k\leq\ind(i)-\ind(i-1)$, hence $O( (\ind(i)-\ind(i-1))\log n)$ time overall.
Taking the sum over all $i=2,\dots,m$, the total time is $O(n\log n)$. 
\qed
\bigskip

\noindent
{\em Proof of Theorem~\ref{thm:size}}.
Since endpoints of all monotone polygonal paths of $\core(C)$ are fixed in minimization problem before Theorem~\ref{thm:size}, we separately consider every corresponding monotone subcloud $C'$ of points (say) $p_1,\dots,p_n$ ordered by their orthogonal projections to the line through the line segment $[p_1p_n]$.
Let $1=\opt(1)<\dots<\opt(k)=n$ be indices of an optimal $\ep$-approximation (polygonal path) $Q$ to $C'$.
In the notations of Lemma~\ref{lem:algorithm} for the approximation error $2\ep$ we will prove below that $\opt(i)\leq\ind(i)$ by induction on $i$.
Then $n=\opt(k)\leq\ind(k)$ and the size $m$ of the list $1=\ind(1)<\dots<\ind(m)=n$, which is found in Lemma~\ref{lem:algorithm}, is at most $k$ as required.
Taking the sum of upper bounds over all monotone paths of $\core(C)$, we conclude that the $2\ep$-Approximate Skeleton $\sk(C)$ has the total number of vertices not greater than that that number for an $\ep$-optimal skeleton $S$.
\smallskip

The base $i=1$ means that $\opt(1)=1=\ind(1)$, i.e. both paths start from the point $p_1$.
In the inductive step assume that $\opt(i-1)\leq\ind(i-1)$.
If $\opt(i)\leq\ind(i-1)$, then $\opt(i)\leq\ind(i)$ and the inductive step is complete.
The remaining case is $\ind(i-1)<\opt(i)$.
Since $Q$ is an $\ep$-approximation to $C'$, we have $d([p_{\opt(i-1)} p_{\opt(i)}], C)\leq\ep$.
Lemma~\ref{lem:segment_error} implies that $d([p_{\ind(i-1)} p_l], C)\leq 2\ep$ for any index $l$ such that $\ind(i-1)<l\leq\opt(i)$.
Lemma~\ref{lem:algorithm}(b) for the approximation $2\ep$ says that $d( [p_{\ind(i-1)} p_{\ind(i)+1}], C )>2\ep$, hence $\opt(i)\leq\ind(i)$.
\qed
\medskip

\begin{dfn}[expanion constants]
\label{dfn:expansion_constant}
Let $C\subset\R^m$ be a cloud and $\bar B(p;r) = \{q \in\R^m \mid d(p,q) \leq r\}$ be the closed ball with the center $p$ and radius $r$. 
The {\em expansion} constant $c_e$ is the smallest real number $c\geq 2$ such that $\forall x : |\bar B(x,2r)| < c|\bar B(x,r)|$.
Let $c_s$ be the similarly defined constant for the metric space of line segments of $\MST(C)$, then set $c=\max\{c_e,c_s\}$.
Other constants $c_p,c_l$ are similarly defined in \cite{beygelzimer2006cover}.
\end{dfn}
\medskip

\noindent
{\em Proof of Corollary~\ref{cor:time}}.
The distance from Definition~\ref{dfn:distances} measured orthogonally to the straight line through fixed endpoints $[p_1 p_n]$ is not smaller than the Hausdorff distance used for $\ep$-offsets in Definition~\ref{dfn:graph}.
Hence the algorithm from Lemma~\ref{lem:algorithm} produces required $\ep$-approximations in the sense of Definition~\ref{dfn:graph}.
The current implementation uses the single-edge clustering based on $\MST(C)$, so Step~1a runs in $O(n)$ time.
The total time is dominated by Step 1b computing $\MST(C)$ in time $O(\max\{c_e^6,c_p^2c^2_l) \}c_e^{10}n\log n\,\al(n))$, where $\al(n)$ is the inverse Ackermann function.
\smallskip

Algorithm~\ref{alg:depths} in Step 1c has the pseudo-code above and maintains a binary tree on $O(n)$ vertices, which requires $O(n\log n)$ time.
Selecting deep vertices in Step~1d and finding longest paths in Step~1e within subtrees of $\MST(C)$ needs $O(n\log n)$ time by classical algorithms \cite{FT87}.
Step~2a to split $C$ into subclouds is implemented by cover trees for line segments of $\core(C)\subset\MST(C)$ in time $O(c_s^{16} n\log n)$ as proved in \cite{beygelzimer2006cover}.
By Lemma~\ref{lem:algorithm} Steps 2b-2g for approximating any subcloud of $n_i$ points by a polygonal path runs in $O(n_i\log n_i)$ time.
Hence the total time at Stage 2 for computing $\sk(C)$ over the cloud $C$ of $n$ points is $O(\max\{c^6,c_p^2c^2_l) \}c^{10}n\log n\,\al(n))$.
\qed

\section{Appendix B: more qualitative comparisons of 3 algorithms}

Fig.~\ref{fig:4star_outputs}--\ref{fig:7star_outputs} show example outputs of 3 algorithms on real and randomly generated clouds in $\R^3$.
In almost all cases the Mapper and $\al$-Reeb graphs contain superfluous short edges, which affect the homeomorphism types.
\smallskip

\begin{figure*}[h]
\includegraphics[width=0.3\textwidth]{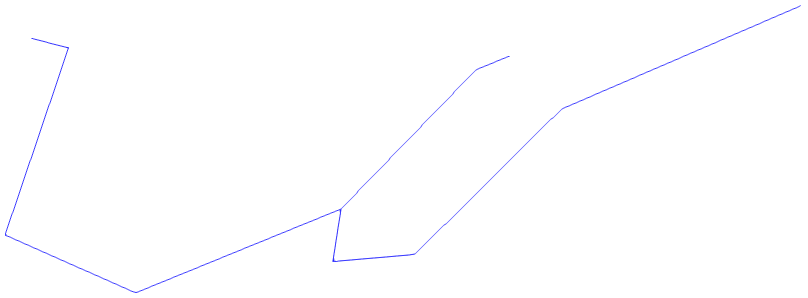}
\hspace*{2mm}
\includegraphics[width=0.3\textwidth]{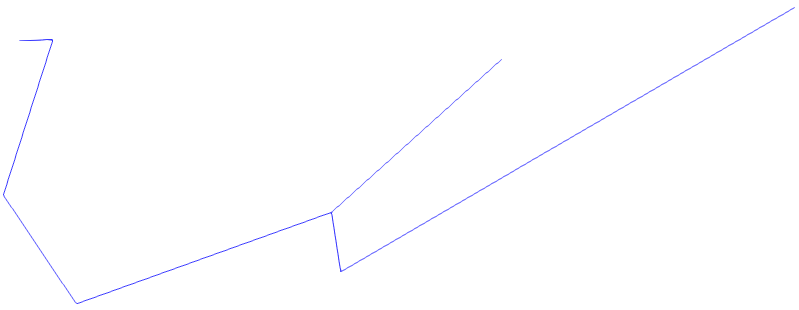}
\hspace*{2mm}
\includegraphics[width=0.3\textwidth]{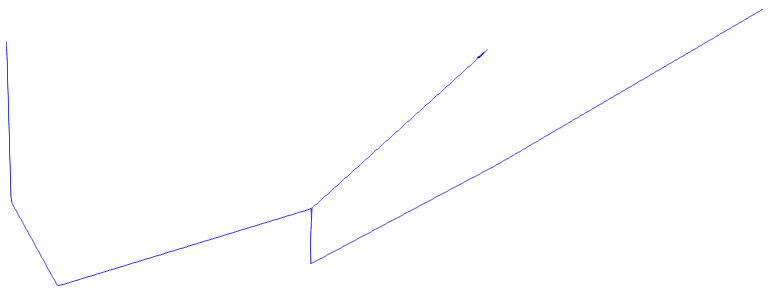}
\caption{\textbf{Left}: $\sk(C)$ for the branching factor $\ga=1.2$. 
\textbf{Middle}: $\ga=1.4$, 
\textbf{Right}: $\ga=1.6$.}
\label{fig:branched_error_varied}
\end{figure*}

\begin{figure*}[h]
\includegraphics[width=0.24\textwidth]{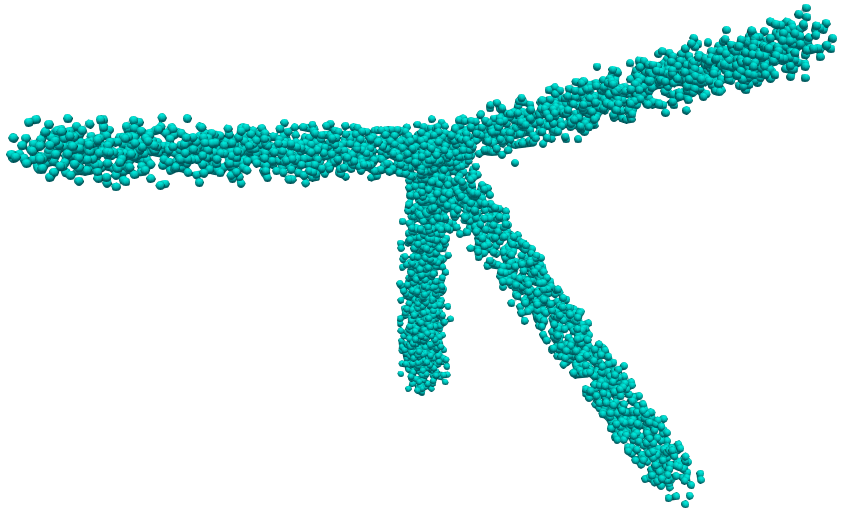}
\includegraphics[width=0.24\textwidth]{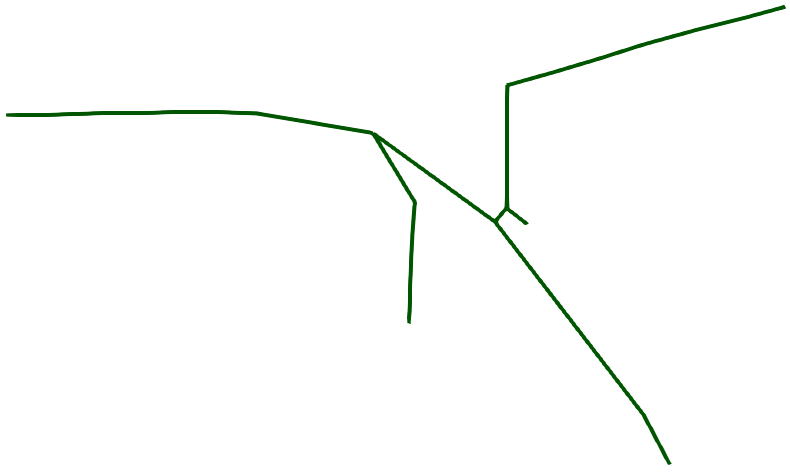}
\includegraphics[width=0.24\textwidth]{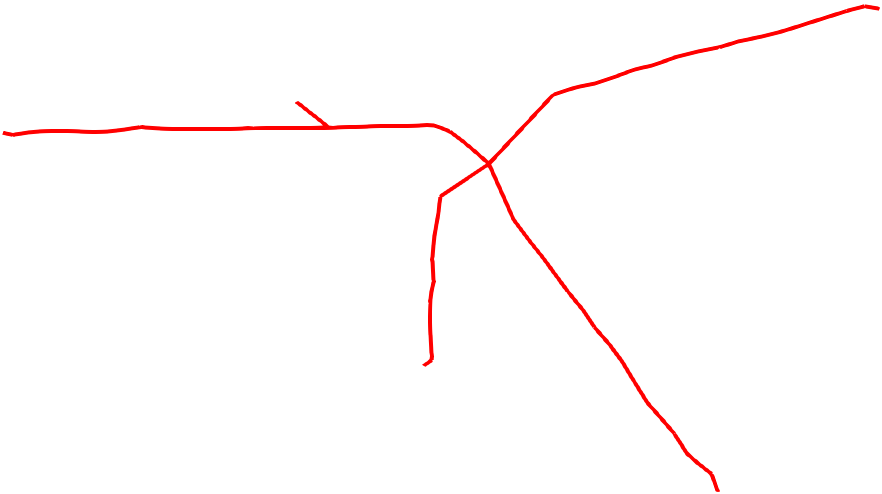}
\includegraphics[width=0.24\textwidth]{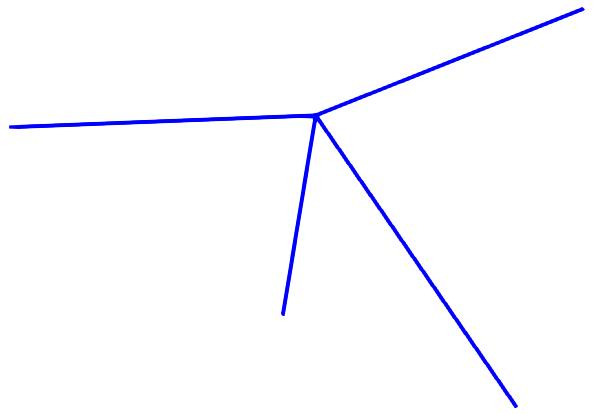}
\caption{\textbf{1st}: a sample around a 4-star in $\R^3$, 
\textbf{2nd}: Mapper, 
\textbf{3rd}: $\al$-Reeb,
\textbf{4th}: $\sk(C)$.}
\label{fig:4star_outputs}
\end{figure*}

\begin{figure*}[h]
\includegraphics[width=0.24\textwidth]{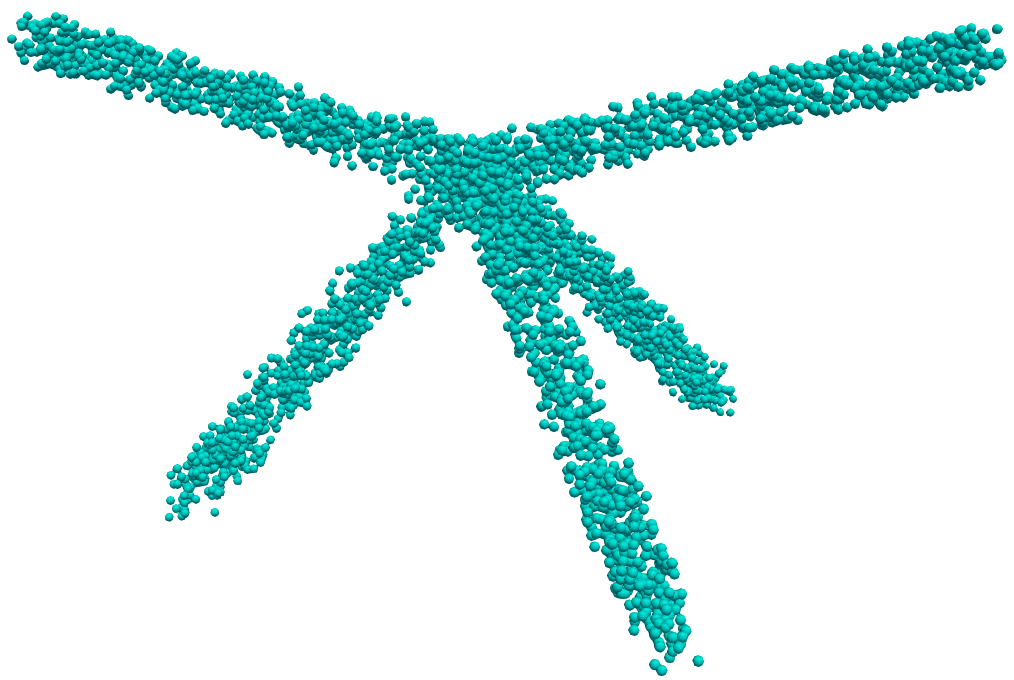}
\includegraphics[width=0.24\textwidth]{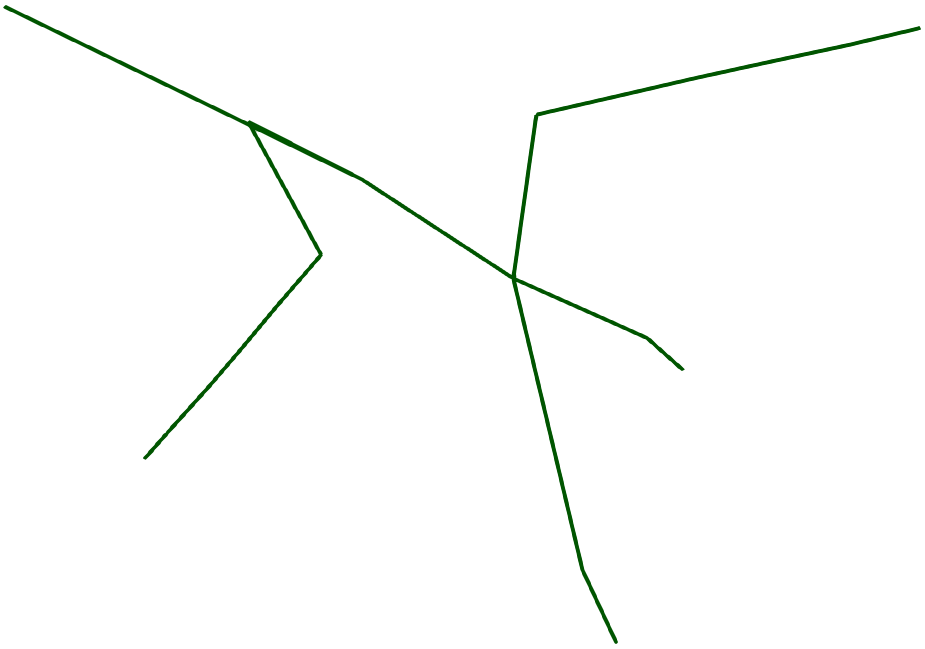}
\includegraphics[width=0.24\textwidth]{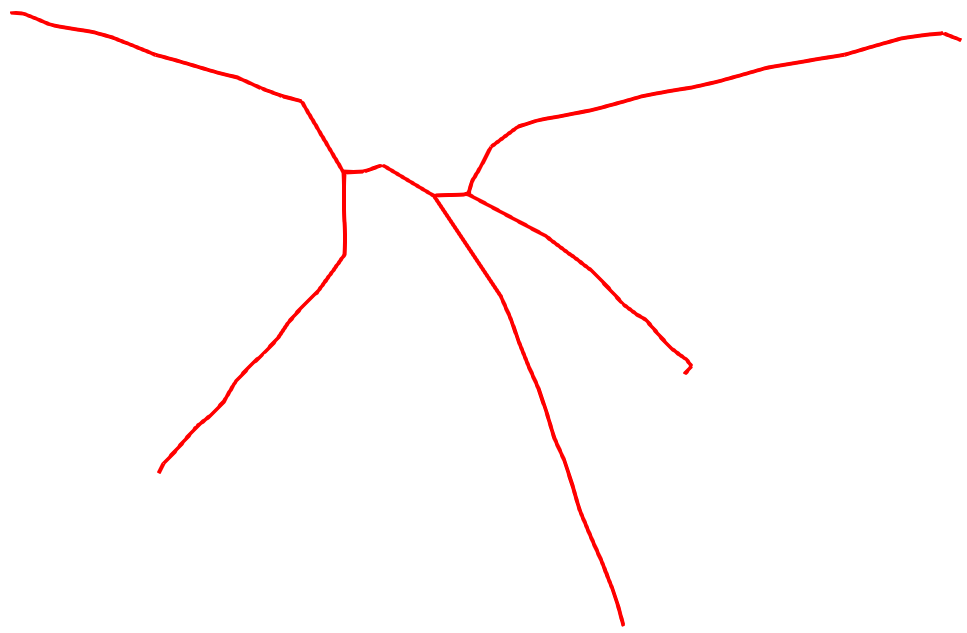}
\includegraphics[width=0.24\textwidth]{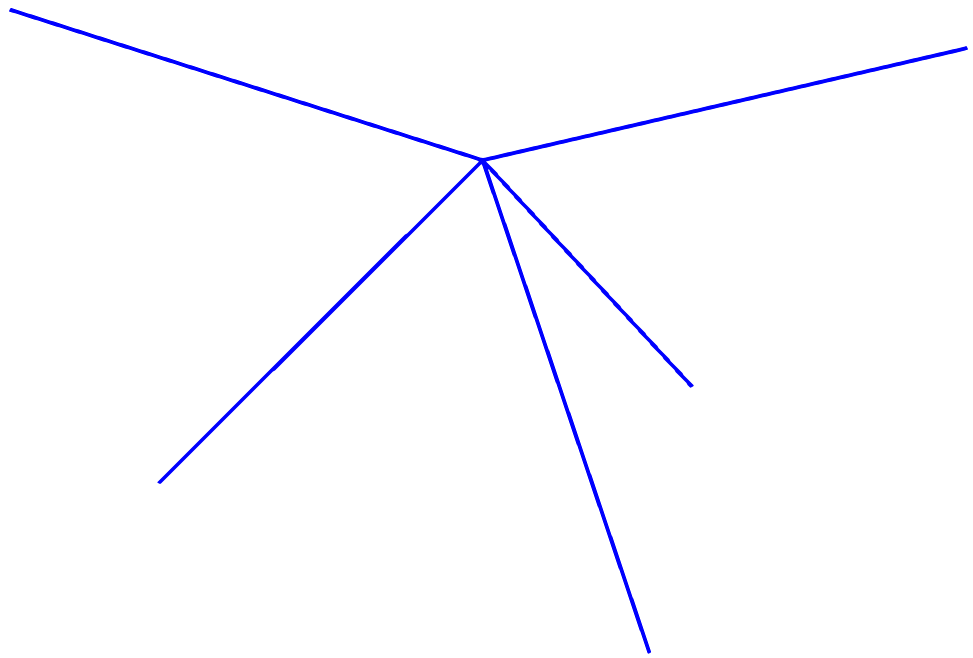}
\caption{\textbf{1st}: a sample around a 5-star in $\R^3$, 
\textbf{2nd}: Mapper, 
\textbf{3rd}: $\al$-Reeb,
\textbf{4th}: $\sk(C)$.}
\label{fig:5star_outputs}
\end{figure*}

\begin{figure*}[h]
\includegraphics[width=0.24\textwidth]{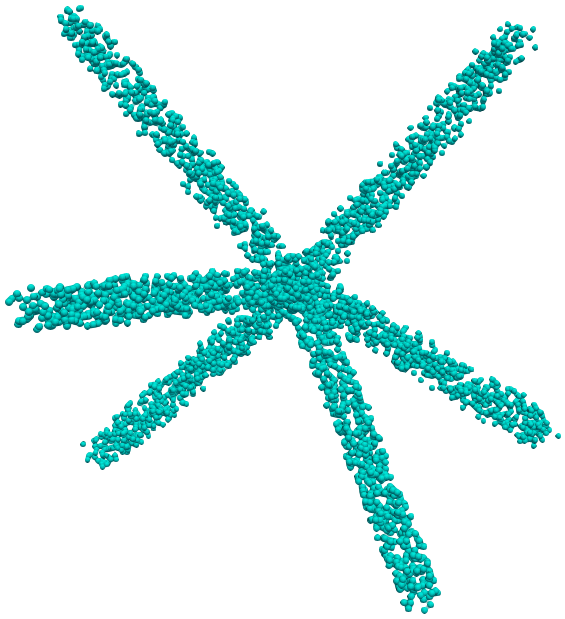}
\includegraphics[width=0.24\textwidth]{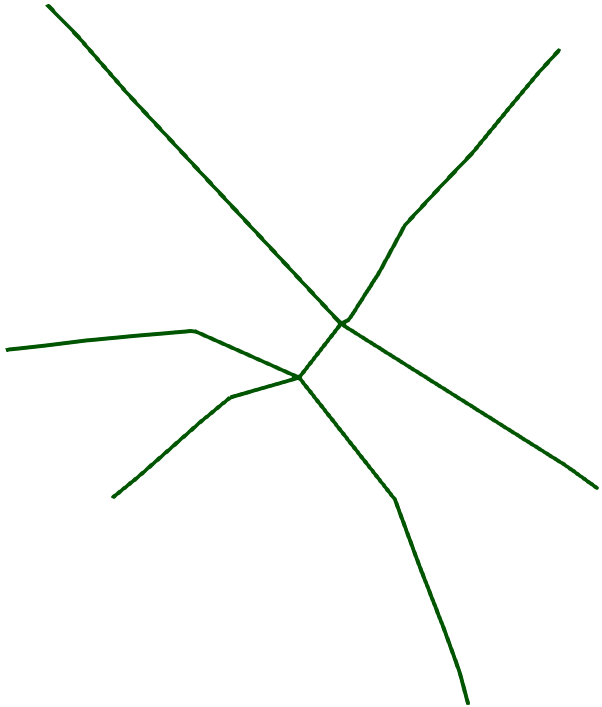}
\includegraphics[width=0.24\textwidth]{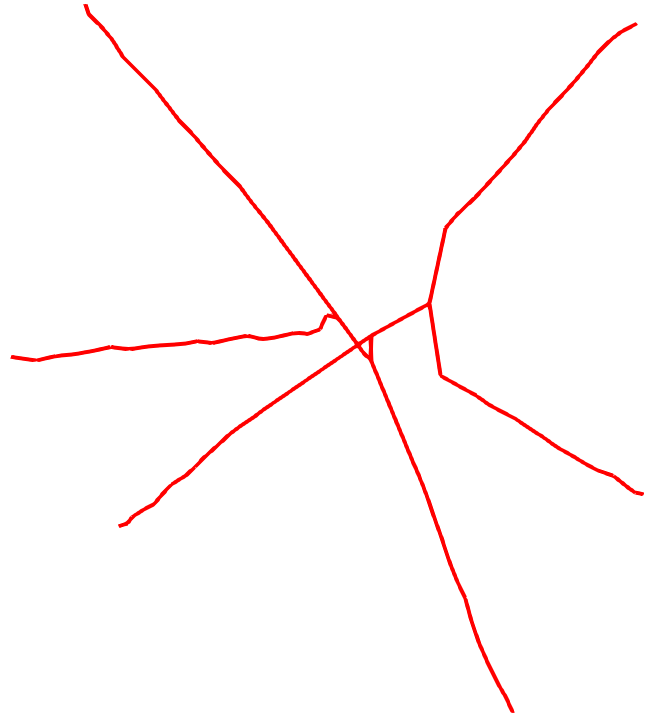}
\includegraphics[width=0.24\textwidth]{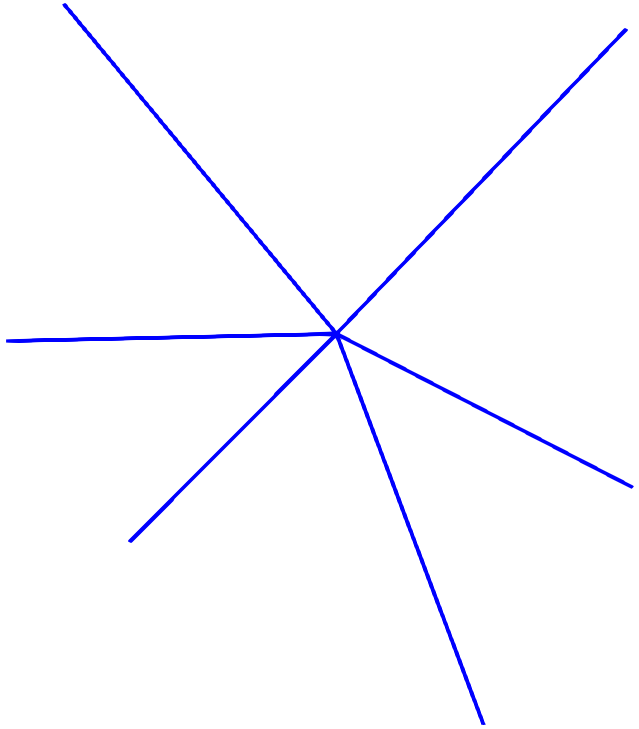}
\caption{\textbf{1st}: a sample around a 6-star in $\R^3$, 
\textbf{2nd}: Mapper, 
\textbf{3rd}: $\al$-Reeb,
\textbf{4th}: $\sk(C)$.}
\label{fig:6star_outputs}
\end{figure*}

\begin{figure*}[h!]
\includegraphics[width=0.24\textwidth]{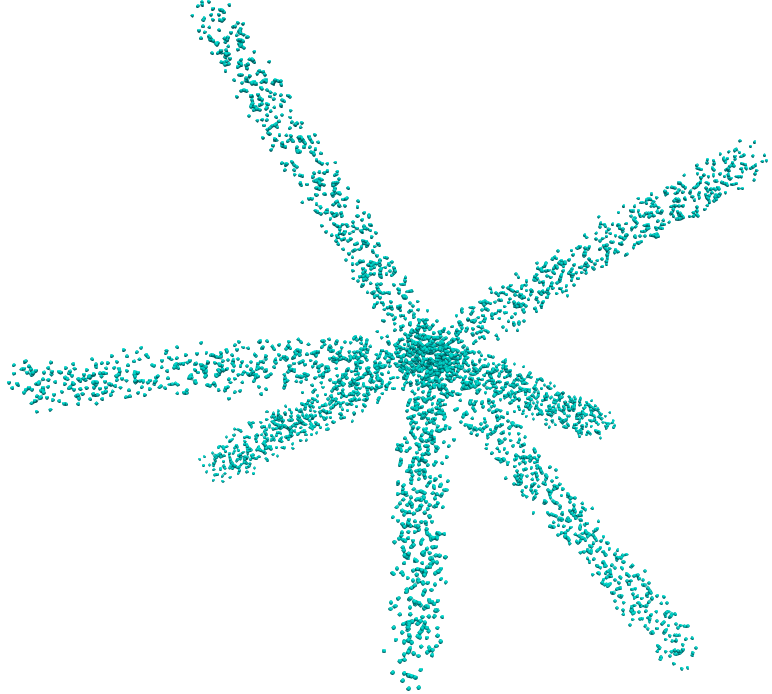}
\includegraphics[width=0.24\textwidth]{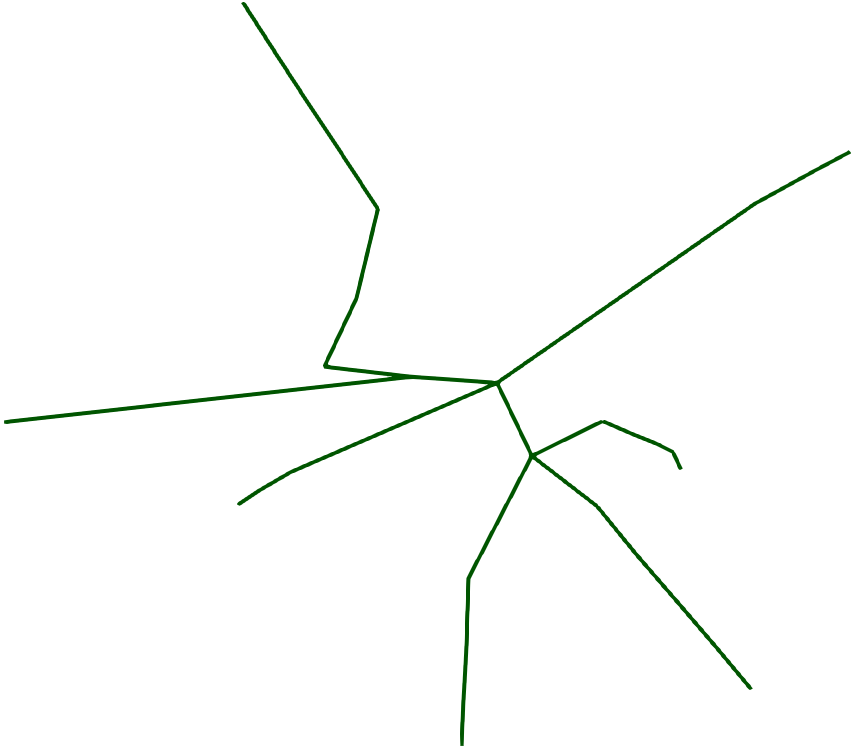}
\includegraphics[width=0.24\textwidth]{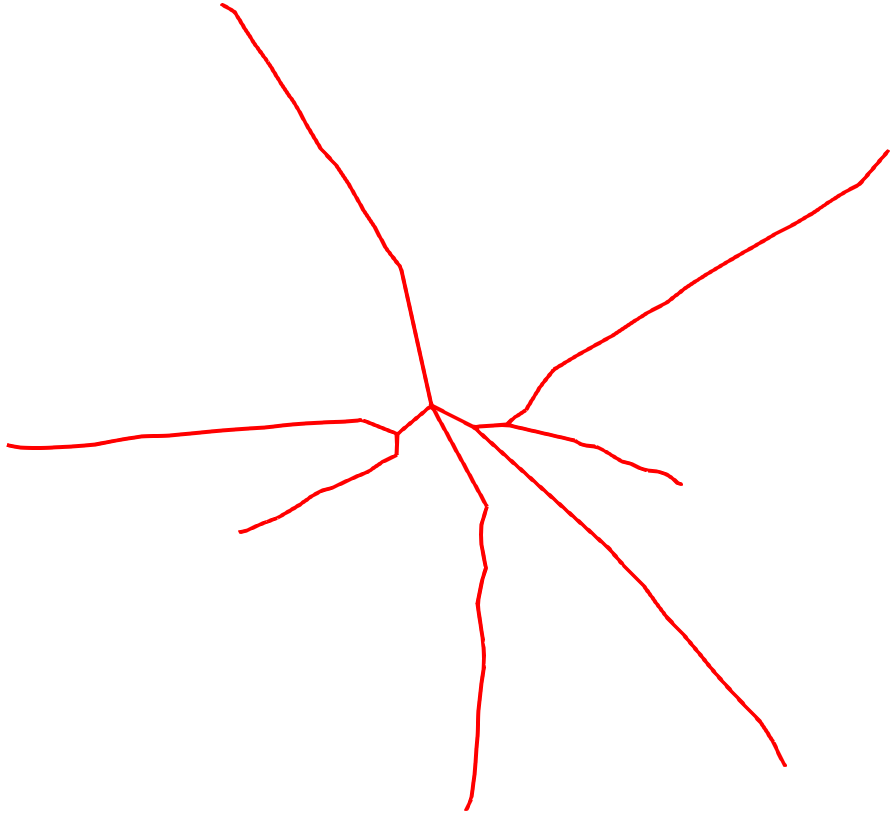}
\includegraphics[width=0.24\textwidth]{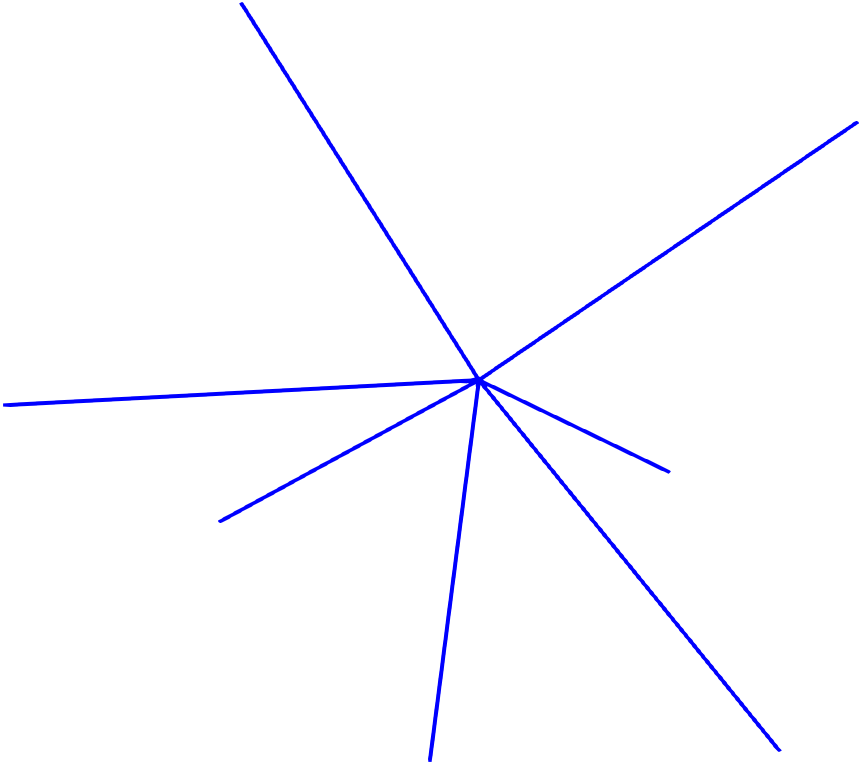}
\caption{\textbf{1st}: a sample around a 7-star in $\R^3$, 
\textbf{2nd}: Mapper, 
\textbf{3rd}: $\al$-Reeb,
\textbf{4th}: $\sk(C)$.}
\label{fig:7star_outputs}
\end{figure*}

The error factor $\ga$ from Corollary~\ref{cor:time} affects the quality of approximation.
Fig.~\ref{fig:branched_error_varied} shows that higher values of $\ga$ lead to more straightened curves.

\begin{figure*}[h!]
\includegraphics[height=21mm]{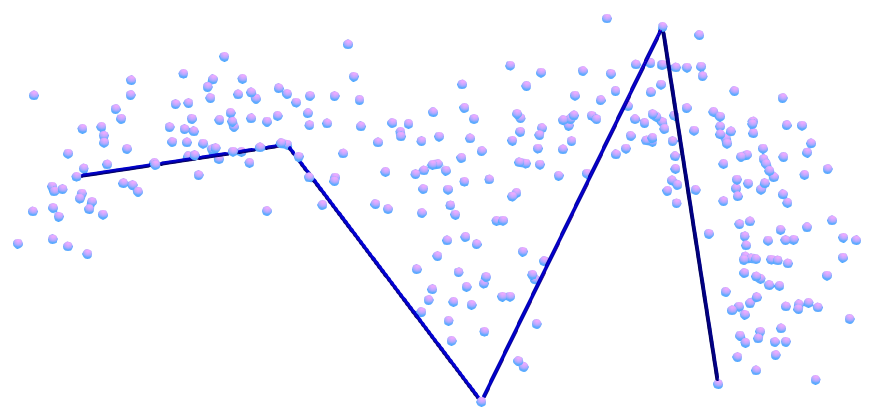}
\hspace*{12mm}
\includegraphics[height=21mm]{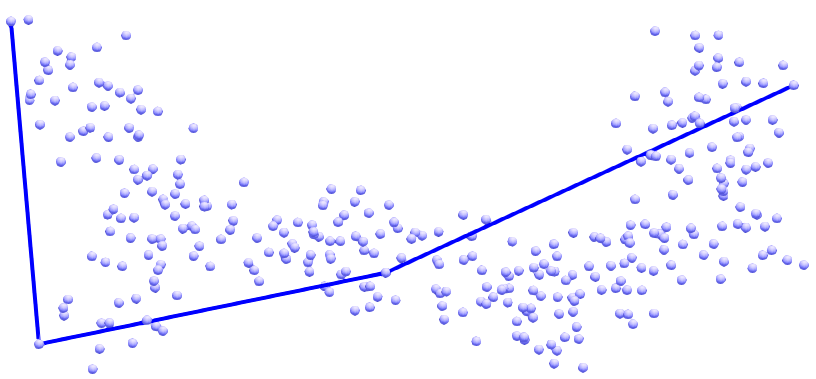}
\caption{Explaining 1.87\% failures of $\sk(C)$ in Table~\ref{tab:micelles}: two micelles $C$ with short edges.
}
\label{fig:hard_micelles}
\end{figure*}

\end{document}